 \documentclass[aps, prd, preprint, preprintnumbers, showpacs, amsfonts, nofootinbib, floatfix]{revtex4}
\usepackage{graphicx}
\usepackage{color}

\def\fourApp{$pp\to \gamma\gamma\gamma\gamma$ }
\def\twoApp{$pp\to \gamma\gamma gg$ }
\def\fourAgg{$gg\to \gamma\gamma\gamma\gamma$ }
\def\twoAgg{$gg\to \gamma\gamma gg$ }
\def\fourAqq{$q{\overline q}\to \gamma\gamma\gamma\gamma$ }
\def\twoAqq{$q{\overline q}\to \gamma\gamma gg$ }

\def\fourAppb{$p{\overline p}\to \gamma\gamma\gamma\gamma$ }
\def\twoAppb{$p{\overline p}\to \gamma\gamma gg$ }

\begin{document}

\preprint{CUMQ-HEP-154}
\title{Collider Effects of Unparticle Interactions in Multiphoton Signals }

\author{T.M. Aliev$^{1}$} 
\email[]{aliev@metu.edu.tr}
\author{Mariana Frank$^2$}
\email[]{mfrank@alcor.concordia.ca}
\author{Ismail Turan$^{2,3}$}
\email[]{ituran@physics.carleton.ca}
\affiliation{$^1$Physics Department, Middle East Technical University, 06531 Ankara, Turkey}
\affiliation{$^2$Department of Physics, Concordia University, 
7141 Sherbrooke Street West, Montreal, Quebec, Canada H4B 1R6.} 
\affiliation{$^3$Ottawa-Carleton Institute of Physics, Carleton University, 1125 Colonel By Drive, Ottawa, Ontario, Canada K1S 5B6.}
\date{\today}
\begin{abstract}
A new model of physics, with a hidden conformal sector which manifests itself as an unparticle  coupling to Standard Model particles effectively through higher dimensional operators, predicts strong collider signals due to unparticle self-interactions. We perform a complete analysis of the most spectacular of these signals at the hadron collider, $pp(\overline{p}) \to \gamma \gamma\gamma \gamma$ and $\gamma \gamma gg$. These processes can go through  the three-point unparticle self interactions as well as through some $s$ and $t$ channel diagrams with one and/or two unparticle exchanges. We  study the contributions of individual diagrams classified with respect to the number of unparticle exchanges and discuss their effect on the cross sections at the Tevatron and the LHC. We also restrict the Tevatron bound on the unknown coefficient of the three-point unparticle correlator. With the availability of data from Tevatron, and the advent of the data emerging from the LHC, these interactions can provide a clear and strong indication of unparticle physics and distinguish this model from other beyond the standard model scenarios.
\end{abstract}
\maketitle
\section{Introduction}\label{sec:intro}
The Large Hadron Collider (LHC) at CERN, once functional, will attempt to discover the last piece of the Standard Model (SM) puzzle (the Higgs boson, responsible for electro-weak breaking) as well as hopefully provide signals of physics beyond the standard model (BSM). The main process responsible for producing a Higgs boson at the LHC is  gluon fusion, and a light boson produced in such fashion is then expected to decay into two photons. Such a decay is supposed to be weak, as Higgs interact only weakly and in the SM the process occurs at one-loop level only. 

In a BSM scenario, a new scalar field which is singlet under the SM gauge group can mix with the Higgs boson but also couple to photons and gluons directly through higher dimensional operators with a cutoff  scale $\Lambda_{\cal U}$. Such a model has been proposed by Georgi \cite{Georgi:2007ek}, based on the hypothesis that there could be an exact scale-invariant hidden sector at a high energy scale. Below $\Lambda_{\cal U}$, the model emerges as interpolating fields with general non-integral scaling dimension $d_{\cal U}$, which behave like a $d_{\cal U}$ number of invisible massless particles--hence the term unparticle ($\cal U$) used by Georgi. An unparticle does not posses a fixed invariant mass squared, but has a continuum mass spectrum, as a consequence of scale invariance
\begin{equation}
\displaystyle
\rho (p^2)=A_{d_{\cal U}}\ \theta (p^0)\ \theta (p^2)\ (p^2)^{d_{\cal U}-2},
\end{equation}
with $A_{d_{\cal U}}$ the normalization factor. 
The production of unparticles at low energies is described by an effective field theory and can give rise to peculiar energy distributions because of the non-integral values of $d_{\cal U}$. 
 
 While the concept dates only to 2007  \cite{Georgi:2007ek}, there have been over 120 studies so far for implications of unparticle physics topics as varied as on neutrino physics \cite{neutrinos} cosmology \cite{cosmology},   rare processes in the standard model \cite{rare}, effects on precision measurements \cite{effects}, as well as specific low energy signals for the detection of unparticles \cite{signals}. The peculiarities of unparticle physics have motivated numerous studies of their implications at the colliders \cite{Chen:2007qr,CKY-short}.
Recently, it has been pointed out \cite{Feng:2008ae, Georgi:2009xq} that among collider signals, unparticle self-interactions give rise to spectacular effects in gluon fusion processes. (Note also the more recent study on unparticle self interactions \cite{Bergstrom:2009iz}). In particular, while the  interactions $gg \to {\cal U} \to \gamma \gamma$ lead to enhanced signals in the Higgs boson decay channels, the three-point self-interactions $gg \to {\cal U} \to {\cal U} {\cal U} \to \gamma \gamma \gamma \gamma$  are practically background free  and may provide an unsuppressed and extremely promising signal for unparticle discovery. The unusual feature here is that the production of additional unparticles with high $p_T$ in the final states does not suppress the production rate, unlike the production of known particles \cite{Feng:2008ae}. There is encouraging  data from the L3 collaboration  on $\gamma \gamma $ interaction, which  exceeds the prediction of the standard model by about one order of magnitude at the highest transverse momentum data \cite{Achard:2001zb},  and gluon fusion might show similar enhancements. Motivated by these considerations, in the present paper we extend the calculation in \cite{Feng:2008ae} and we present a complete evaluation of the relevant diagrams, in a comparative fashion, for  the processes $pp(\overline{p}) \to \gamma \gamma\gamma \gamma$ and $\gamma \gamma gg$ at both the Tevatron and the LHC. We restrict our analysis to the hadron colliders: the LHC because of the large center-of-mass energy, and the Tevatron, because it operates now and could produce visible signals of physics BSM. A similar analysis is possible for ILC, but for our purposes, it is beyond the scope of this work. We also update the Tevatron bound on the three-point unparticle self interaction discussed in \cite{Feng:2008ae}, based on the complete evaluation of the cross section.

We proceed as follows: in Section \ref{sec:model} we describe briefly the unparticle model features, concentrating on the ones relevant for our considerations. In Section \ref{sec:calculation} we present the numerical investigation of processes generated by unparticle self interactions in $ \gamma \gamma\gamma \gamma$ and $ \gamma \gamma gg$ at both the Tevatron and the LHC. We compare the results obtained from production subprocesses, grouped by number of unparticles in intermediate states, and by channels. There are basically only two important unknown parameters, namely the scaling dimension $d_{\cal U}$ and the cutoff scale $\Lambda_{\cal U}$. We restrict a third parameter, emerging from the three-point unparticle correlation function from the Tevatron data. We discuss the results obtained and conclude in Section \ref{sec:conclusion}. In the Appendix we give the complete list of relevant diagrams for our calculation.

\section{Unparticle Formulation}\label{sec:model}

 In this section, we summarize the basic features of unparticle physics. In order to be relevant, unparticles must interact with SM fields, although the  detailed dynamics of the interactions are not known.
 These interactions are parameterized in the effective Lagrangian approach as  interactions between a new physics operator ${\cal O} _{\cal U}$ with dimension $d_{\cal U}$ and the standard model operator ${\cal O} _{SM}$ with dimension $d_{SM}$ 
 \begin{equation}
\displaystyle
 {\cal L}=\lambda_{d_{SM}}\ \Lambda_{\cal U}^{4-d_{\cal U}-d_{SM}}\ {\cal O}_{\cal U}\ {\cal O}_{SM}
 \end{equation}
with $\lambda_{d_{SM}}$ a coupling constant of order 1. From scale invariance considerations, the phase space  for the unparticle operator is a  continuous spectral density
\begin{equation}
\displaystyle
d{\rm LISP}_{d_{\cal U}}=A_{d_{\cal U}}\ \theta (p^0)\ \theta (p^2)\ (p^2)^{d_{\cal U}-2} \frac{d^4 p}{(2 \pi)^4}
\end{equation}
where $A_{d_{\cal U}}$ is the normalization factor 
\begin{equation}
\displaystyle
A_{d_{\cal U}}= \frac{16 \pi^{\frac52}}{(2 \pi)^{2 d_{\cal U}}}\frac{\Gamma\left ( d_{\cal U}+ \frac12 \right)}{\Gamma\left ( d_{\cal U}-1 \right)\Gamma\left ( 2d_{\cal U} \right)}
\end{equation}
Unparticle operators interact with the SM fields through the exchange
of heavy particles of mass $M$. Integrating out the heavy fields produces a series of effective 
operators describing unparticle interactions with the SM fields at low energy \cite{Georgi:2007ek}.    
The operators describing the interactions for various unparticle spins are \cite{Georgi:2007ek,Chen:2007qr}
\begin{eqnarray}
\displaystyle
\mathrm{Spin \; 0:} &&
\lambda_0^\prime  \frac{1}{\Lambda_{U}^{d_U - 1}} \overline f f {\cal O}\;, \;\; 
\lambda_0^{\prime\prime} \frac{1}{\Lambda_{U}^{d_U-1} } \overline f i \gamma^5 f {\cal O}\;, \;\; 
\lambda_0 \frac{1}{\Lambda_{U}^{d_U }} G_{\alpha\beta} G^{\alpha\beta} {\cal O} \;, \; \cdots
\nonumber \\
\mathrm{Spin \; 1:} &&\lambda_1 \frac{1}{\Lambda_{U}^{d_U - 1} }\, \overline f \gamma_\mu f \,
{\cal O}^\mu \;, \;\;
\lambda_1^\prime \frac{1}{\Lambda_{U}^{d_U - 1} }\, \overline f \gamma_\mu \gamma_5 f \,
{\cal O}^\mu \;, \; \cdots
\nonumber \\
\mathrm{Spin \; 2:} && 
- \frac{1}{4}\lambda_2^\prime \frac{1}{\Lambda_{U}^{d_U} } \overline \psi \,i
   \left(  \gamma_\mu \stackrel{\leftrightarrow}{{\bf D}}_\nu + 
           \gamma_\nu \stackrel{\leftrightarrow}{{\bf D}}_\mu \right )
  \psi  \,  {\cal O}^{\mu\nu} \; , 
  \lambda_2 \frac{1}{\Lambda_{U}^{d_U} } G_{\mu\alpha} 
G_{\nu}^{\;\alpha} {\cal O}^{\mu\nu} \; , \; \cdots
\nonumber 
\end{eqnarray}
For the spin-0 case, the interaction of $\cal U$ with a fermion pair is suppressed if there is a $\gamma_5$ in the vertex and we neglect such terms. Note also that, unlike the above form, in our study this vertex is defined with a Higgs coupling such that the prefactor becomes $e \lambda_0^\prime v/\Lambda_{\cal U}^{d_{\cal U}}$ where $v$ is the Higgs vacuum expectation value. Otherwise, one needs to introduce two cutoff scale. 

The propagators for the unparticles were suggested to be   \cite{CKY-short}:
\begin{equation}
\displaystyle
\Delta_F(p^2)=\frac{A_{d_{\cal U}}}{2 \sin \pi d_{\cal U}}\frac{i}{(|p^2|+i 0^+)^{2-d_{\cal U}}}\ e^{i\,\phi}
\end{equation}
for the scalar, and
\begin{equation}
\displaystyle
\left[\Delta_F(p^2)\right]_{\mu \nu}=\frac{A_{d_{\cal U}}}{2 \sin \pi d_{\cal U}}\frac{i}{(|p^2|+i0^+)^{2-d_{\cal U}}}\ e^{i\,\phi}\, \pi_{\mu \nu}(p)
\end{equation}
and 
\begin{equation}
\displaystyle
\left[\Delta_F(p^2)\right]_{\mu \nu, \rho \sigma}=\frac{A_{d_{\cal U}}}{2 \sin \pi d_{\cal U}}\frac{i}{(|p^2|+i0^+)^{2-d_{\cal U}}}\ e^{i\,\phi}\, T_{\mu \nu, \rho \sigma}(p)
\end{equation}
for the spin-1 and spin-2 unparticle, respectively. The phase is defined as $\phi\equiv Arg\left[(-p^2)^{d_{\cal U}}\right]$. Thus, it is non-zero only in the s-channel where $p^2$ is positive while $\phi$ is identical to zero in the $t-$ and $u-$ channels. Of course, the $+i0^+$ piece is only relevant in the $s$ channel. Here $\displaystyle \pi_{\mu \nu}=-g_{\mu \nu} +\frac {p_\mu p_\nu}{p^2}$ and $T_{\mu \nu, \rho \sigma}=\frac12\{ \pi_{\mu \rho}(p)\pi_{\nu \sigma}(p) +\pi_{\mu \sigma}(p)\pi_{\nu \rho}(p)-\frac23 \pi_{\mu \nu}(p)\pi_{\rho \sigma}(p)\}$.

For a scalar operator, unitarity requires as a lower bound, $d_{\cal U} \ge 1$. While no upper bound exists, supersymmetric examples suggest values of $d_{\cal U} <2$ \cite{Zhang:2007ih}. For vector unparticle operators $d_{\cal U} >3$ and for symmetric and antisymmetric tensor
operators $d_{\cal U} >2$ and $d_{\cal U} >4$ respectively \cite{Grinstein:2008qk, Mack:1975je}.
 The cross section is dependent on $d_{\cal U}$ as $1/\Lambda^{2d_{\cal U}}$ and it becomes much smaller for
the values of $d_{\cal U}$ larger than three. For this reason in this work we only consider the scalar unparticle effects.

A scalar unparticle couples to the Standard Model sector through:
\begin{equation}
{\cal L}_{\cal U}=-\frac{\lambda_g}{4}\frac {\cal U}{\Lambda^{d_{\cal U}}}G_{\mu \nu}^A G^{A\, \mu \nu}-\frac{\lambda_\gamma}{4}\frac {\cal U}{\Lambda^{d_{\cal U}}}F_{\mu \nu} F^{\mu \nu},
\end{equation}
The Feynman rules for the scalar and tensor unparticle operators coupling to the $gg$ and $\gamma \gamma$ are, respectively
\begin{eqnarray}
{\cal O}:~~~&&4i \lambda^0_{g, \gamma}\frac {\cal U}{\Lambda^{d_{\cal U}}} \left (-p_1 \cdot p_2 g^{\mu \nu}+ p_1^\nu p^\mu\right )\nonumber \\
{\cal O}^{\mu \nu}:~~~&& \lambda^2_{g, \gamma} \frac{ {\cal U}_{\rho \sigma}}{\Lambda^{d_{\cal U}}} [K^{\mu \nu \rho \sigma} + K^{\mu \nu\sigma\rho }]
\end{eqnarray}
where $K^{\mu \nu \rho \sigma}= -g^{\mu \nu}p_1^\rho p_2^\sigma - g^{\rho \mu}g^{\sigma \nu} p_1 \cdot p_2 +g^{\sigma \mu }p_1^\nu p_2^\rho$, and $ \lambda^0,  \lambda^2$ are the scalar and tensor coupling constants.

Evaluation of the processes $pp(\overline{p}) \to \gamma \gamma gg, \, \gamma \gamma \gamma \gamma$ require evaluation of both the two point correlation function \cite{Georgi:2007ek,CKY-short}:
\begin{eqnarray}
\langle 0| {\cal O}_{\cal U}(x) {\cal O}_{\cal U}^{\dagger}(0)|0 \rangle=\int \frac{d^4P}{(2 \pi)^4} e^{-iPx}\rho_{\cal U}(P^2)
\end{eqnarray}
as well as the three point interaction \cite{Feng:2008ae}
\begin{eqnarray}
\displaystyle
\langle 0| {\cal O}_{\cal U}(x) {\cal O}_{\cal U}(y){\cal O}_{\cal U}^{\dagger}(0)|0 \rangle=\frac{C_d^{\prime}}{(|x-y|\ |x|\ |y|)^{d_{\cal U}}}
\end{eqnarray}
The later  yields, in momentum space, using the two point correlation function:
\begin{eqnarray}
\displaystyle
&&\langle 0| {\cal O}_{\cal U}(p_1) {\cal O}_{\cal U}(p_2){\cal O}_{\cal U}^{\dagger}(p_1+p_2)|0 \rangle \nonumber \\
&&=
C_d \int\frac{d^4q}{(2\pi)^4}\left \{\left [-q^2-i\epsilon\right] \left [-(p_1-q)^2-i\epsilon\right] \left [-(p_2+q)^2-i\epsilon\right] \right \}^{\frac{d_{\cal U}}{2}-2}\nonumber \\
&&=-i(-1)^nC_d\left (\frac1s \right)^{n-2}F_y\left (\frac{p_1^2}{s}, \frac{p_2^2}{s}\right )
\end{eqnarray}
where $n=6\,(1-d_{\cal U}/4)$ and 
\begin{eqnarray} 
\displaystyle
F_y\left (\frac{p_1^2}{s}, \frac{p_2^2}{s}\right )=\frac{\Gamma(n-2)}{16 \pi^2\left[ \Gamma\left(\frac{n}{3} \right) \right]^3} \int_0^1 dx_1 dx_2 dx_3 (x_1x_2x_3)^{\frac{n}{3} -1}\delta(x_1+x_2+x_3-1)\left( \frac{1}{\Delta}\right)^{n-2}
\end{eqnarray}
with $\Delta =x_1\, x_2\, p_1^2/s+x_1\,x_3\,p_2^2/s+x_2\,x_3$ and $s=(p_1+p_2)^2$. As a check, we reproduced the Fig.~2 in \cite{Feng:2008ae} where $F_y$ is plotted as a function of its one argument and got complete agreement.

In all the analyses that  follow we take $\lambda_{g,\gamma}^0=1$ and $\lambda_0^\prime = \sqrt{2\pi}/e$. We restrict our considerations to the scalar unparticle, as it was shown that the vector and tensor give smaller contributions \cite{CKY-short}\footnote{ Note that the vector unparticle does couples only $q{\overline q}$ while the tensor couple to both $q{\overline q}$ and $gg$.}.

\section{Numerical Analysis}\label{sec:calculation}

The complete set of Feynman diagrams for the subprocesses $gg,\, q{\overline q} \to {\cal U}... \to \gamma \gamma \gamma \gamma$  and for the subprocesses $gg,\, q{\overline q} \to {\cal U}... \to \gamma \gamma gg$ are given in the Appendix. There are quite a number of diagrams contributing to the processes, requiring some automatization to do calculation. We generated our results using the softwares {\tt FeynArts} and {\tt FormCalc} \cite{Hahn:2000jm}  and the {\tt Vegas} Monte Carlo program for the four-body phase space integration as well as for the numerical evaluation of the function $F_y$. In order to avoid collinear singularities, we employ angular cuts as well as cuts on the transverse energies of the final photons and gluons. We require the final photons and gluons to have at least 30 GeV transverse energy and their ejection directions make at least 10 degrees with the beamline in the forward and backward directions. Since the unparticle model has unconventional features,  the vertices and propagators the softwares {\tt FeynArts} and {\tt FormCalc} required some modifications. The hadronization is done by using the {\tt LHAPDF} version 5.3 \cite{Whalley:2005nh} with the parton distribution functions of {\tt CTEQ6LL}. We neglect the masses of the light quarks in the proton while calculating the contributions from the quark pair. We also neglect the contributions from the SM diagrams but include the SM background separately. In fact,  the SM backgrounds for \fourAppb and  the signals \twoApp go only through quark subprocesses at partonic level. Under the same conditions, we estimated the background cross sections as follows: at the Tevatron;  $2.6\times 10^{-5}$ pb and $0.3$ pb for \fourAppb and \twoAppb, respectively; at the LHC, $7.7\times 10^{-5}$ pb (for \fourApp) and $0.8$ pb (for \twoApp).

In order to keep track of various contributions and make comparisons, we group the graphs according to the partonic constituents ($gg$ or $q{\overline q}$), the number of unparticles present in the intermediate states, and whether they proceed through the $s$ or the $t$ channel. We follow the same path in the numerical evaluation of each contribution, which we discuss in the following subsections.

The existence of more than one external gluon requires an extra care. To preserve the gauge invariance one either needs to include the ghost contributions or perform the the gluon polarization sum for only the transversal  polarization. We follow the second path  by introducing appropriate projection operators for the transverse polarization states of the gluon. Since one can still do the sum conventionally for one of the gluons, in the $gg\to \gamma\gamma\gamma\gamma$ and $gg\to \gamma\gamma gg$ subprocesses, we need a single and three projection operators, respectively, when summation over the gluon polarizations is performed (for more details, see \cite{Cutler:1977qm}).

\subsection{The multiphoton processes at Tevatron}

While the LHC seems to be still besieged by technical problems, it is worthwhile to
explore signals coming the physics of the currently running
largest energy collider facility, the Tevatron at Fermilab. Tevatron has accumulated yet unexplored data, and can still expected to yield signals for physics beyond the standard model. With this in mind, we begin the  exploration of unparticle interactions in the multiphoton signals at the Tevatron, specifically looking at the signals \fourAppb and \twoAppb. The center of mass energy $\sqrt{s}=1.96$ GeV for Tevatron is used throughout.

\subsubsection{The process  \fourAppb at Tevatron} 
\begin{figure}[b]
\begin{center}
\hspace*{-1.2cm}
        $\begin{array}{cc}
	\includegraphics[width=3.5in,height=3.2in]{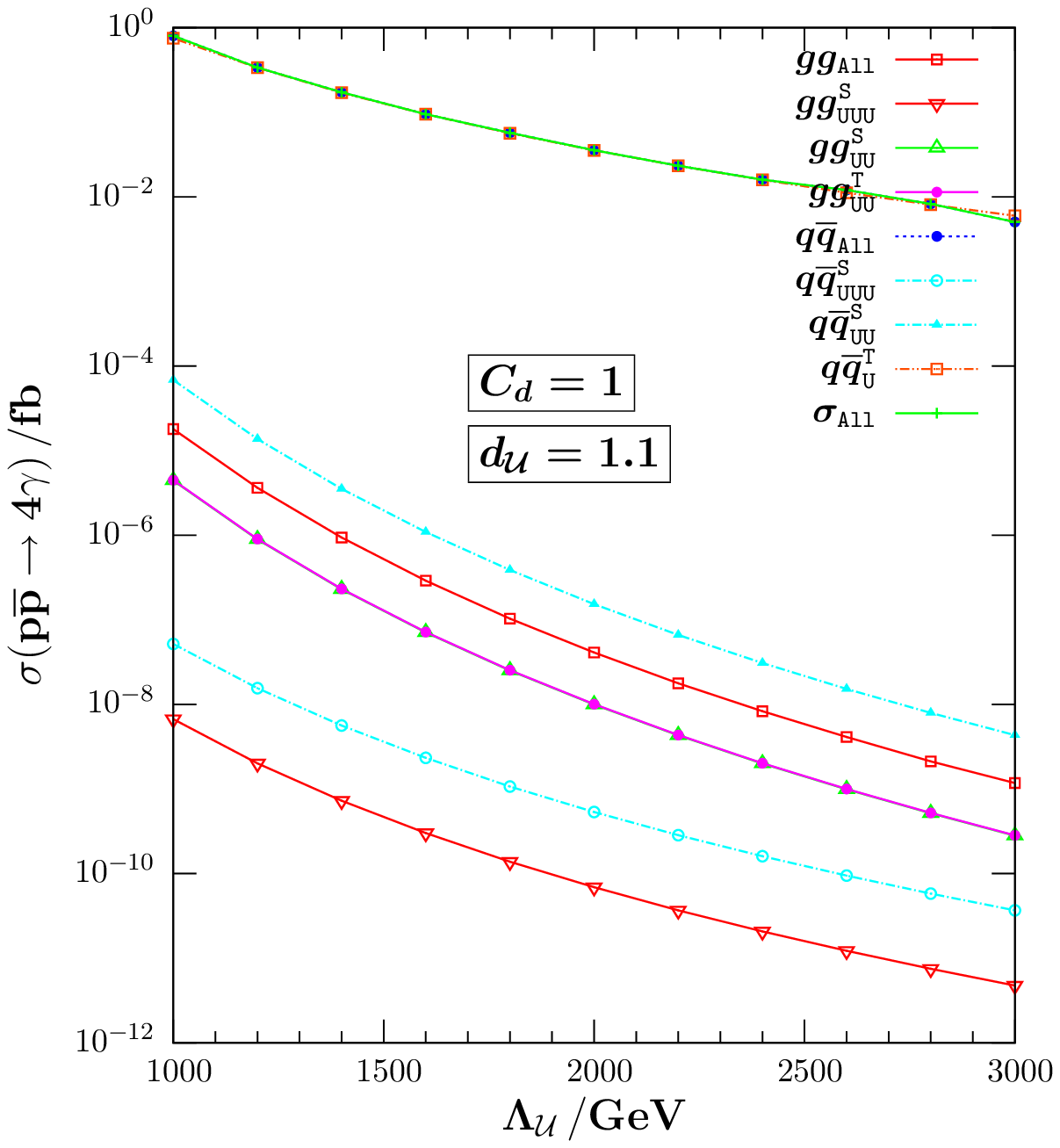} &\hspace*{-0.8cm}
	\includegraphics[width=3.5in,height=3.2in]{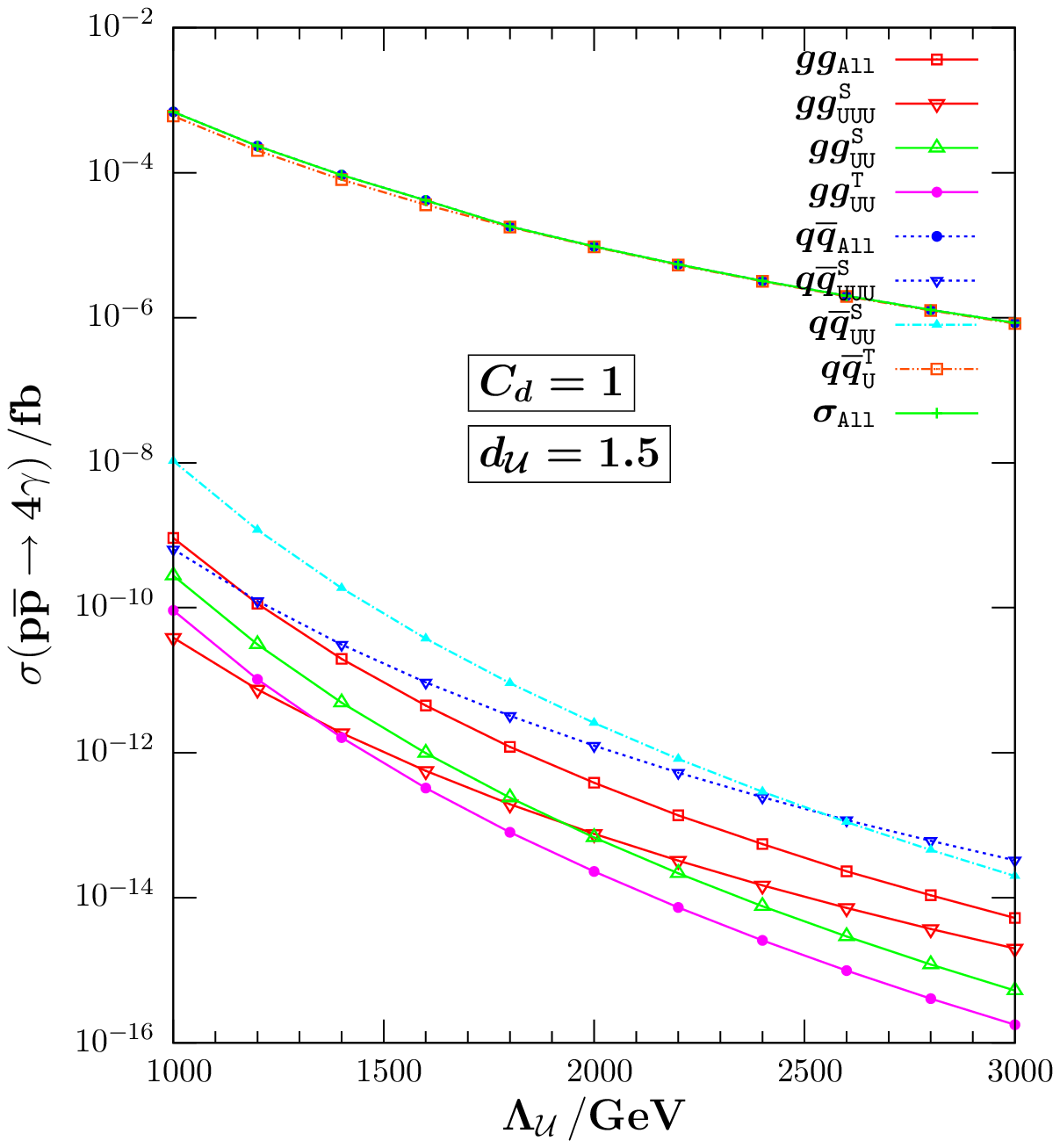}\\
	\includegraphics[width=3.5in,height=3.2in]{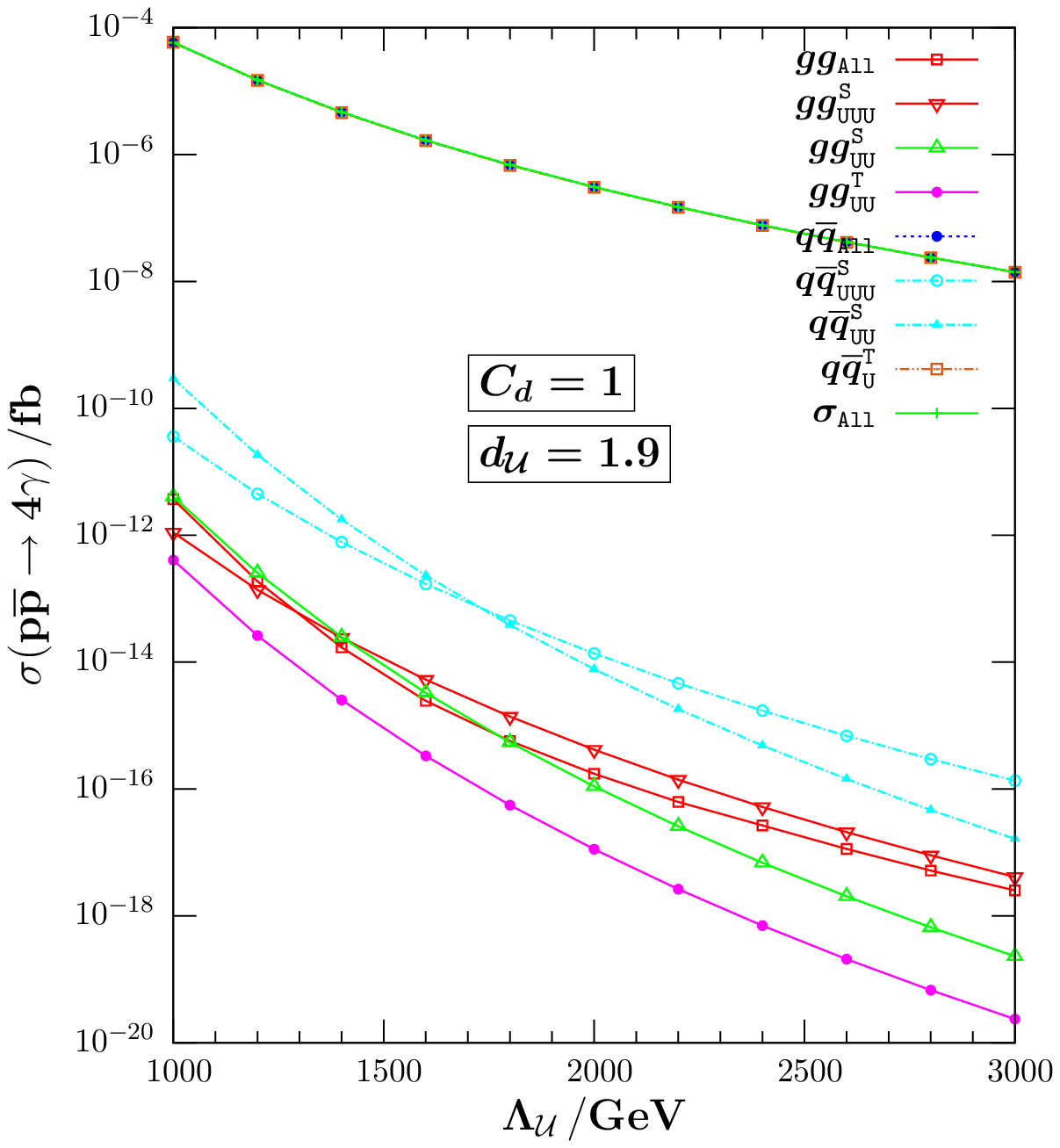}
\end{array}$
\end{center}
\vskip -0.2in
\caption{The \fourAppb  cross section at $\sqrt{s}=1.96$ TeV as a function of $\Lambda_{\cal U}$ for $C_d=1$ and $d_{\cal U}=1.1$, $d_{\cal U}=1.5$ and $d_{\cal U}=1.9$. The individual contributions from the $gg$ and $q\bar{q}$ subprocesses are grouped and shown for each channel and number of unparticles exchanged.}\label{4Ad_tev}
\end{figure}
\begin{figure}[htb]
\begin{center}
\hspace*{-1.2cm}
        $\begin{array}{cc}
	\includegraphics[width=3.5in,height=3.7in]{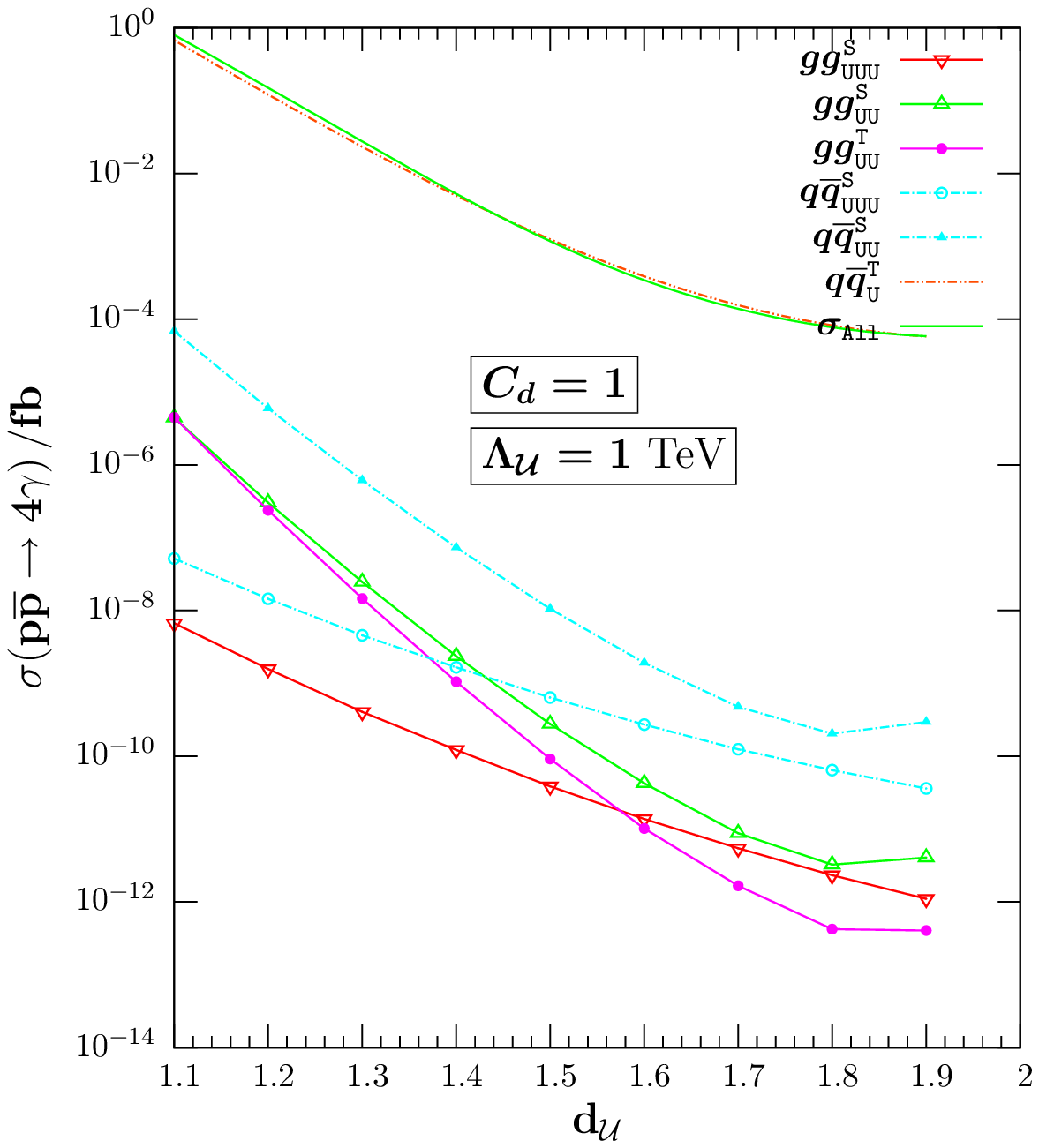} &\hspace*{-0.8cm}
	\includegraphics[width=3.5in,height=3.7in]{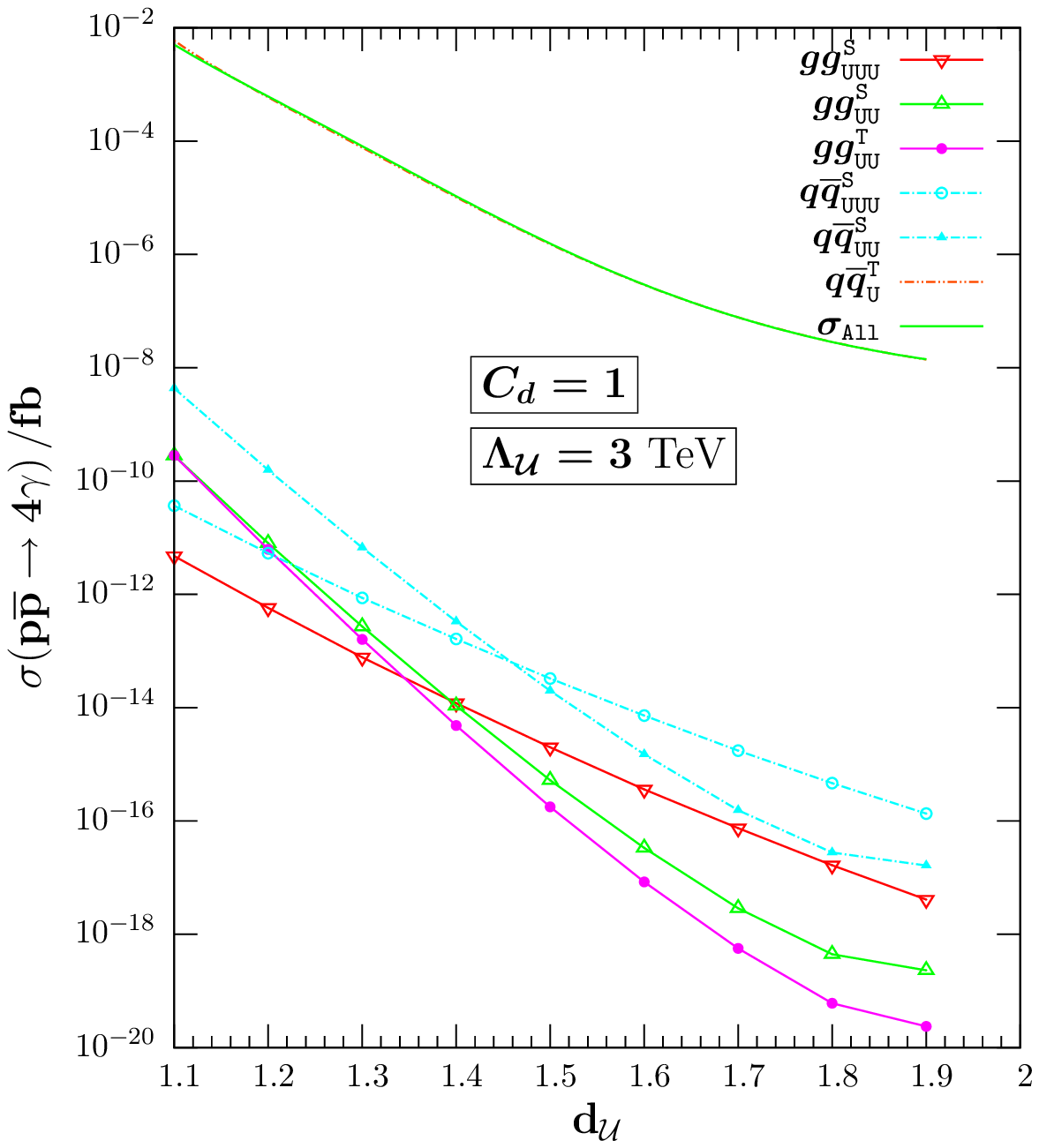}
\end{array}$
\end{center}
\vskip -0.2in
\caption{The \fourAppb  cross section at $\sqrt{s}=1.96$ TeV as a function of $d_{\cal U}$  for $C_d=1$ and $\Lambda_{\cal U}=1$ TeV and $\Lambda_{\cal U}=3$ TeV. The individual contributions from the $gg$ and $q\bar{q}$ subprocesses are grouped and shown for each channel and number of unparticles exchanged.}\label{4Alamda_tev}
\end{figure}
We explore first the dependence of the total cross section on the value of the parameter $\Lambda_{\cal U}$. Fig.~\ref{4Ad_tev} shows the variation of the cross section with the unparticle energy scale for three values of $d_{\cal U}$ (1.1, 1.5 and 1.9). As expected, the cross section is the largest for the smallest $d_{\cal U}$. Within each panel, we  plot separately contributions from the two partonic channels ($gg$ and $q{\overline q}$) and momentum transfer channels ($s$ and $t$), and present separate contributions coming from one, two and three unparticle exchange. One of the processes, the $s$-channel contribution coming from the exchange of three unparticles as intermediate states in $gg$ and $q {\overline q}$ was calculated in \cite{Feng:2008ae}. Clearly, as can be seen from all three graphs in Fig.~\ref{4Ad_tev}, this process is subdominant for all values of $\Lambda_{\cal U}$. As expected, the processes with the least number of unparticles dominate, as we must include in the evaluation one order of the energy scale $\Lambda_{\cal U}$ for every unparticle exchanged. In addition, processes proceeding through the  $s$ channel are suppressed compared to the $t$-channel, and the $gg$ partonic contributions are much smaller at the Tevatron than the $q{\overline q}$ ones. We take $C_d=1$ for our evaluations; however later we will use the known information on the Tevatron data to restrict $C_d$ and discuss its effect to the maximal cross section. Looking at all the contributions, the cross section is dominated by the $q{\overline q}$ partonic contribution with one unparticle exchange proceeding through the $t$-channel, and could reach as much as around $1$~fb. The cross section is very sensitive to the values of the $d_{\cal U}$ parameter, and is maximal for smaller $d_{\cal U}$.

In Fig.~\ref{4Alamda_tev} we show the dependence of the total cross section at the Tevatron on the $d_{\cal U}$ parameter for two fixed values of the unparticle scale $\Lambda_{\cal U}=1,~3$ TeV. As each unparticle intermediate state contributes to the amplitude a factor of $\displaystyle \frac{1}{\Lambda_{\cal U}^{2d_{\cal U}}}$, the drop in cross section with increasing $\Lambda_{\cal U}$ is expected. The  \fourAppb  cross section at $\sqrt{s}=1.96$ TeV is dominated again by the $q{\overline q}$ partonic contribution with one unparticle exchange proceeding through the $t$-channel, and for $\Lambda_{\cal U}=1$ TeV, $d_{\cal U}=1.1$, we recover the maximal ($\approx 1$~fb) value for the cross section. As a common feature of both parameter investigations, the cross section for the process \fourAppb is completely dominated by one set of diagrams only, the $t$ channel $g{\overline q}$ with one unparticle exchange. We should also note that the diagrams with two $\cal U$ exchange in the $t$-channel, given in the panel (c) of Fig.~\ref{fd_qq4A_1} in Appendix, are proportional the mass of the quarks and we neglect them. This is also true for the subprocess \twoAqq.


\subsubsection{The process \twoAppb at Tevatron}

\begin{figure}[htb]
\begin{center}
\hspace*{-1.2cm}
        $\begin{array}{cc}
	\includegraphics[width=3.5in,height=3.2in]{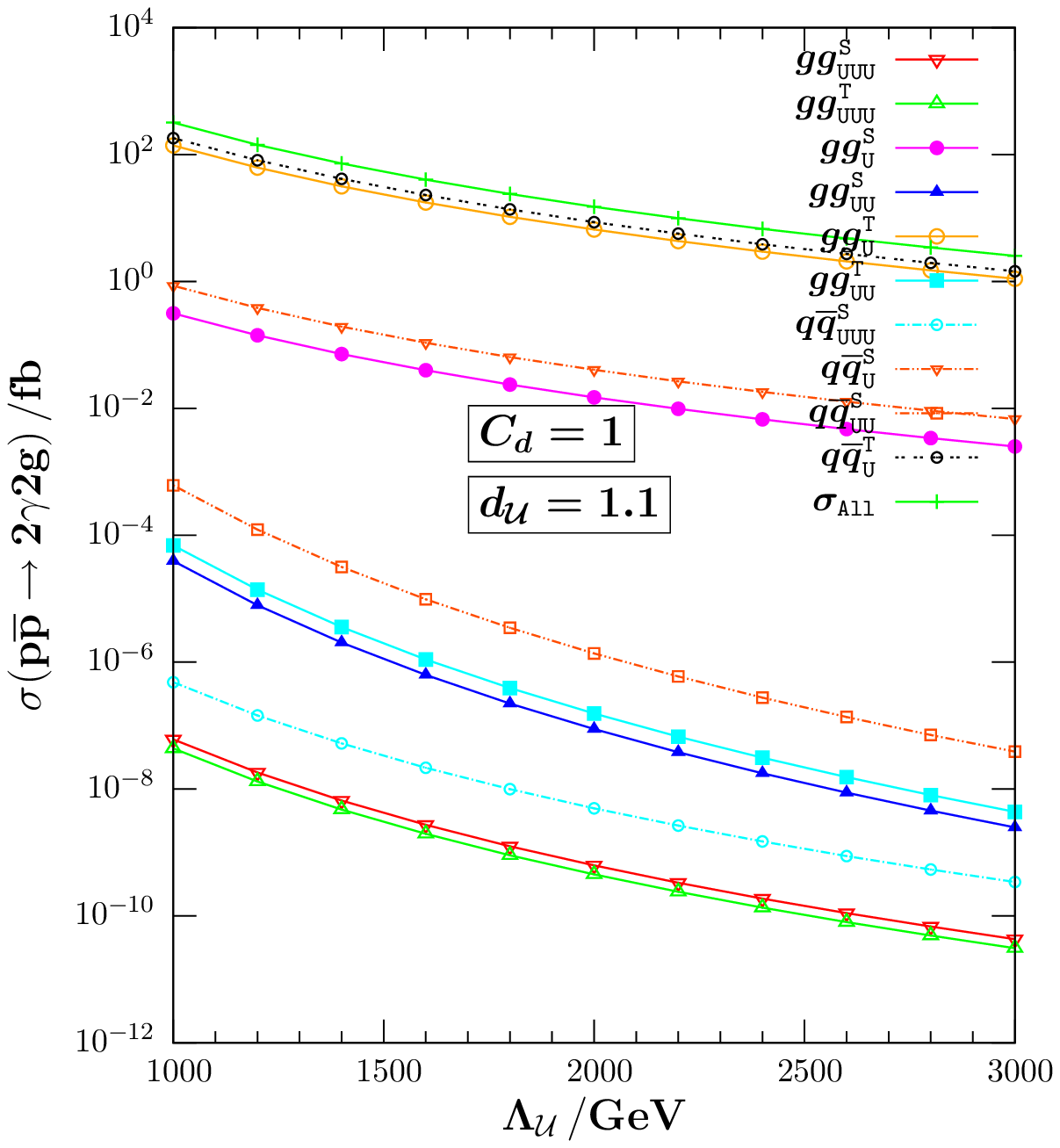} &\hspace*{-0.8cm}
	\includegraphics[width=3.5in,height=3.2in]{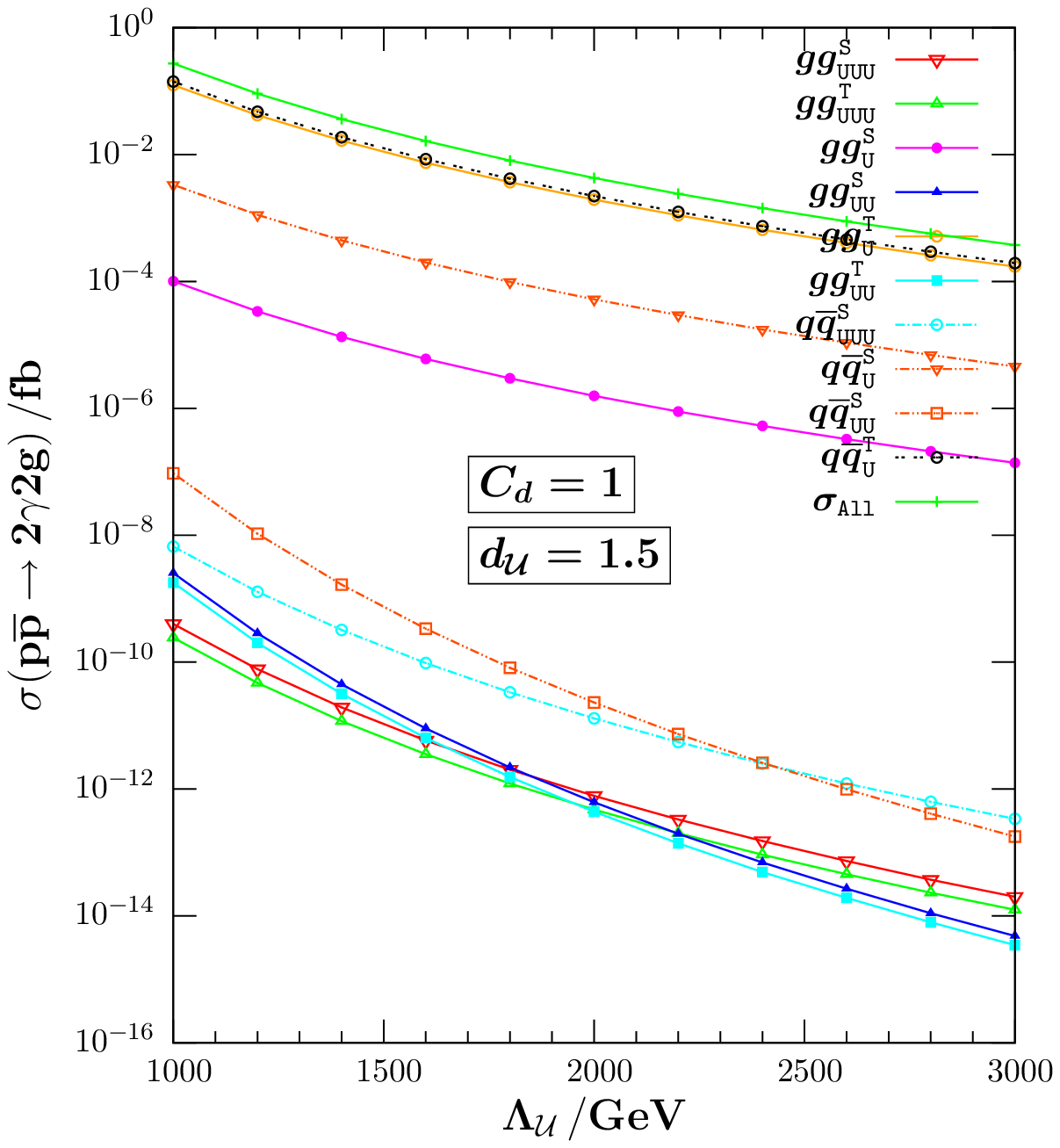}\\
	\includegraphics[width=3.5in,height=3.2in]{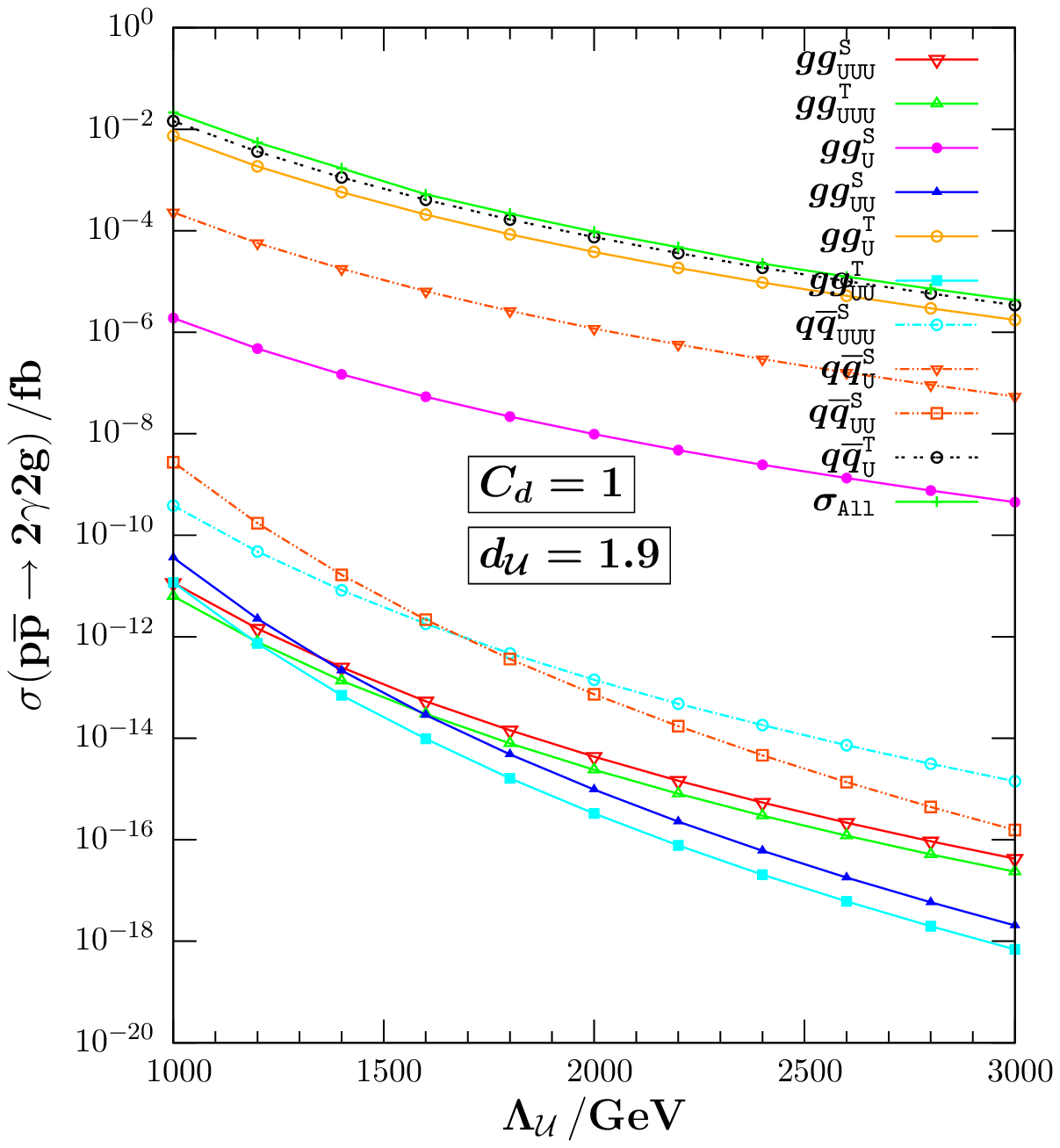}
\end{array}$
\end{center}
\vskip -0.2in
\caption{The \twoAppb  cross section at $\sqrt{s}=1.96$ TeV as a function of $\Lambda_{\cal U}$ for $C_d=1$ and $d_{\cal U}=1.1$, $d_{\cal U}=1.5$ and $d_{\cal U}=1.9$. The individual contributions from the $gg$ and $q\bar{q}$ subprocesses are grouped and shown for each channel and number of unparticles exchanged.}\label{2A2gd_tev}
\end{figure}
\begin{figure}[htb]
\begin{center}
\hspace*{-1.2cm}
        $\begin{array}{cc}
	\includegraphics[width=3.5in,height=3.7in]{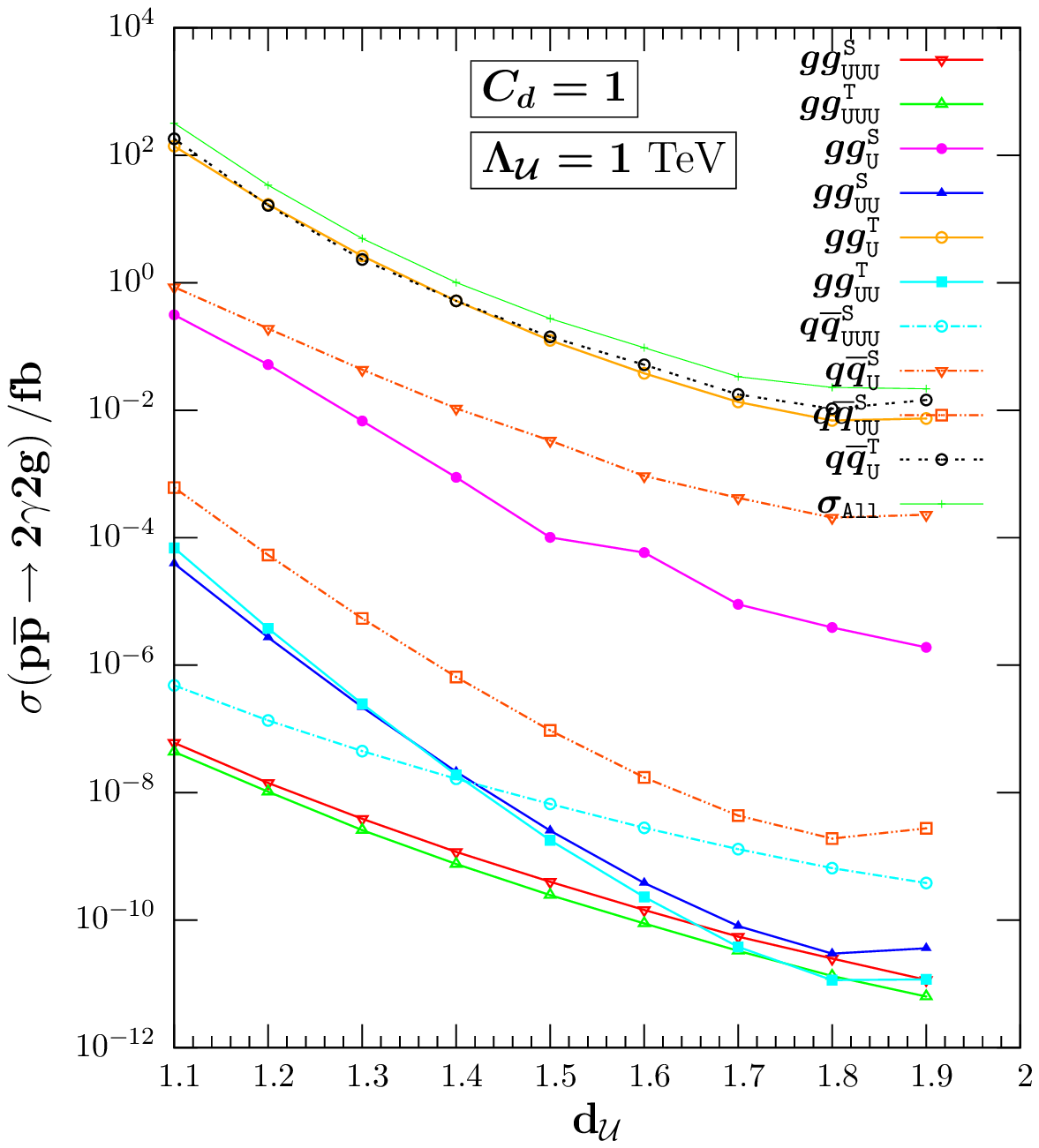} &\hspace*{-0.8cm}
	\includegraphics[width=3.5in,height=3.7in]{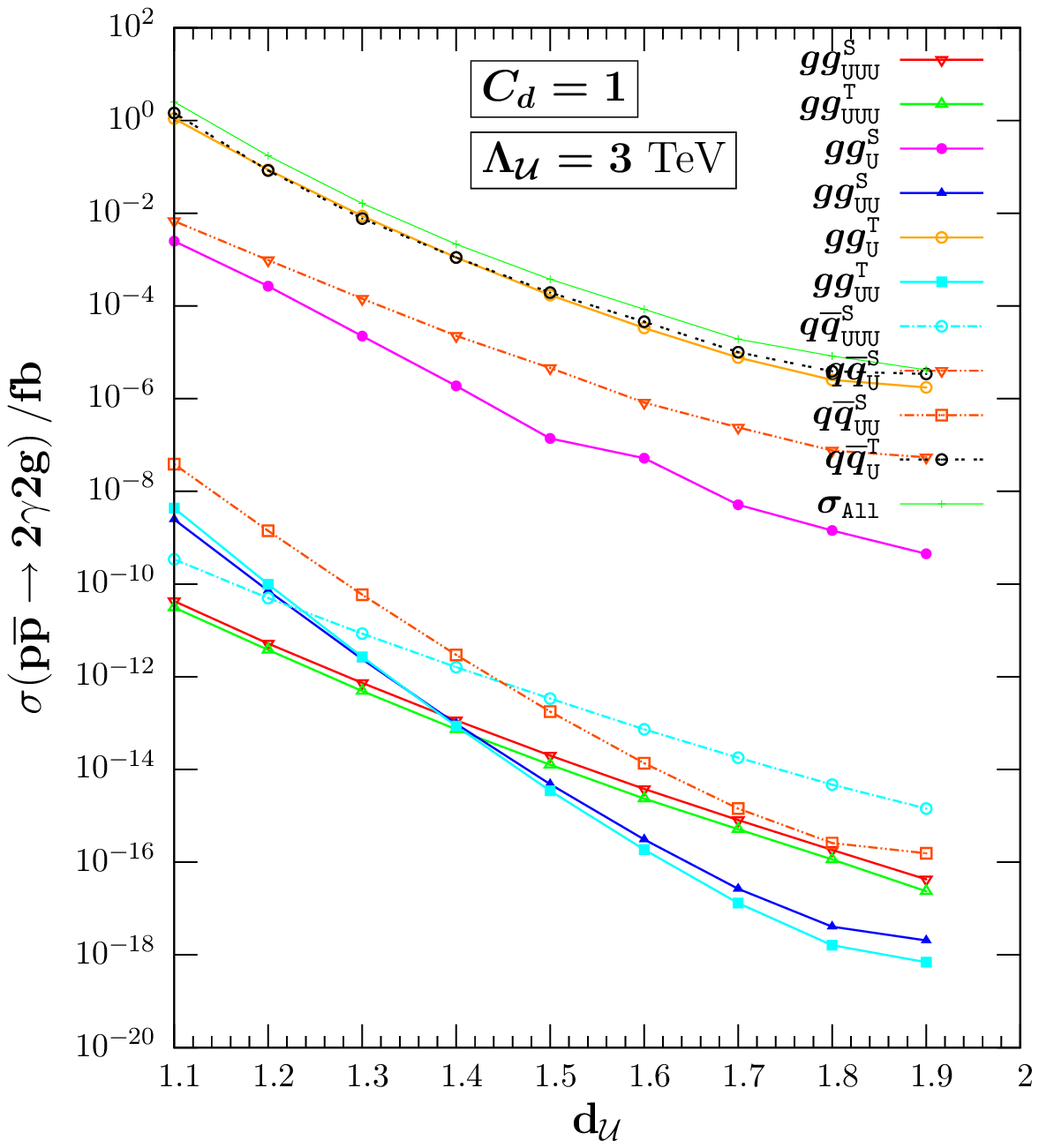}
\end{array}$
\end{center}
\vskip -0.2in
\caption{The \twoAppb  cross section at $\sqrt{s}=1.96$ TeV as a function of $d_{\cal U}$  for $C_d=1$ and $\Lambda_{\cal U}=1$ TeV and $\Lambda_{\cal U}=3$ TeV. The individual contributions from the $gg$ and $q\bar{q}$ subprocesses are grouped and shown for each channel and number of unparticles exchanged.}\label{2A2glamda_tev}
\end{figure}

We perform the same analysis for the process  \twoAppb at the Tevatron. Unlike the four photon signal, this involves two jets which would be affected by background signals and whose identification requires further studies. The Feynman diagrams corresponding to this process are given in the Appendix.  As before, the relevant parameters are $d_{\cal U}$ and $\Lambda_{\cal U}$ and we use the settings as in the four-photon case. In Fig.~\ref{2A2gd_tev} we show the dependence of the total cross section on energy scale of the unparticle, for three values of $d_{\cal U}$ (1.1, 1.5 and 1.9); and in Fig.~\ref{2A2glamda_tev} we give the corresponding dependence on the $d_{\cal U}$ parameter, for two values of $\Lambda_{\cal U}=1,~3$ TeV.
The cross section in this case is dominated by two contributions:  the partonic contributions from $q{\overline q}$ and $gg$ containing the minimal number of unparticle as intermediate states (one in this case), and without $s$-channel suppression (both $t$-channels). The cross section is further enhanced here with respect to \fourAppb by the presence of two powers of the strong coupling constant $\alpha_s$ replacing the electroweak one $\alpha_{w}$, and could reach about $100$ fb for the $\Lambda_{\cal U}=1$ TeV, $d_{\cal U}=1.1$ case. Fig.~\ref{2A2glamda_tev} shows that the maximal value for the cross section decreases by two order of magnitudes when we increase $\Lambda_{\cal U}$ from $1$ to $3$ TeV. It is interesting to note that while $gg$ contribution is suppressed at the Tevatron for hadron production, and in  unparticle-mediated \fourAppb, it is very strong for  the process \twoAppb, while all the other subprocesses are subdominant.

\subsection{The multiphoton processes at LHC}


\begin{figure}[htb]
\begin{center}
\hspace*{-1.2cm}
        $\begin{array}{cc}
	\includegraphics[width=3.5in,height=3.2in]{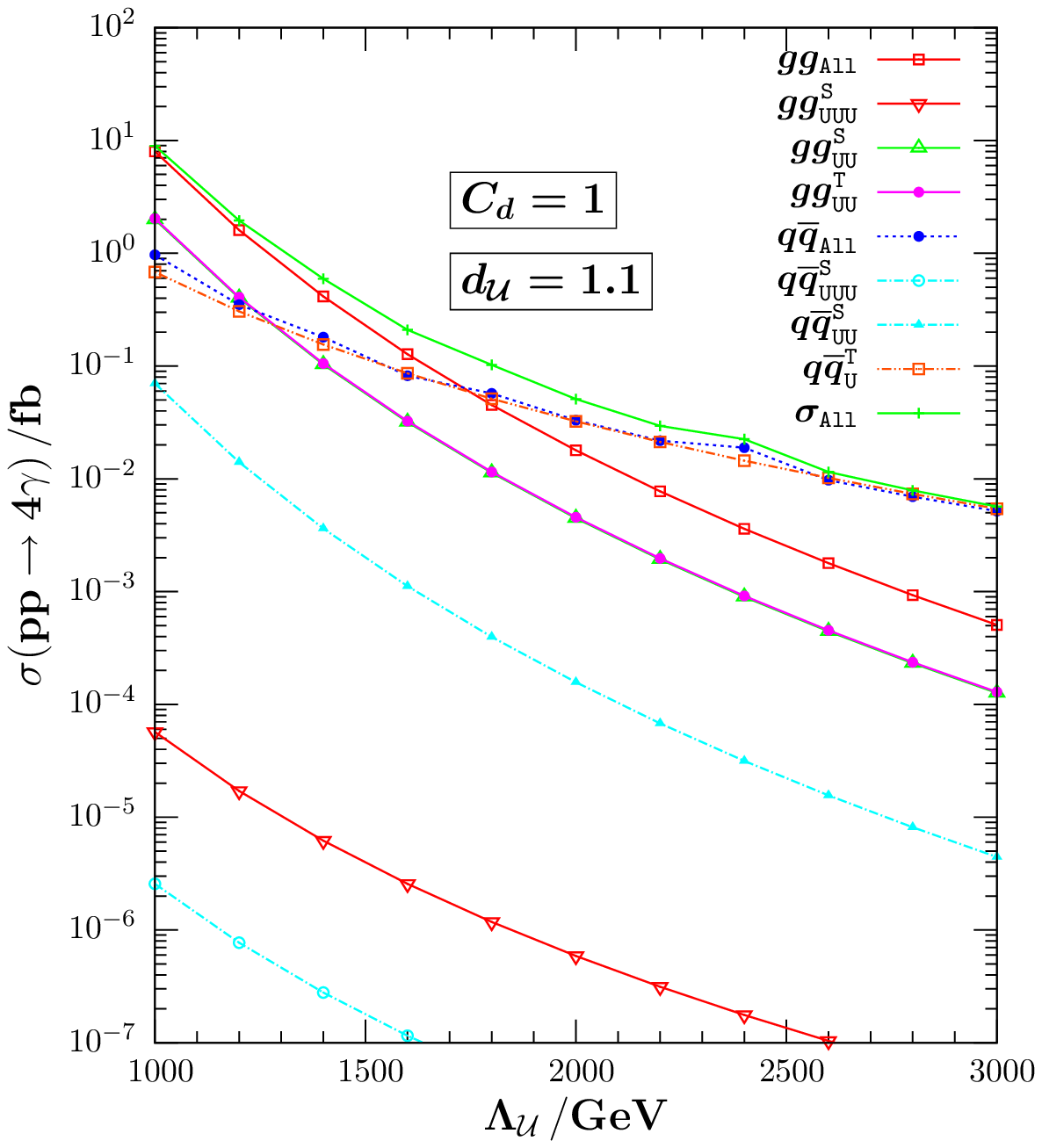} &\hspace*{-0.8cm}
	\includegraphics[width=3.5in,height=3.2in]{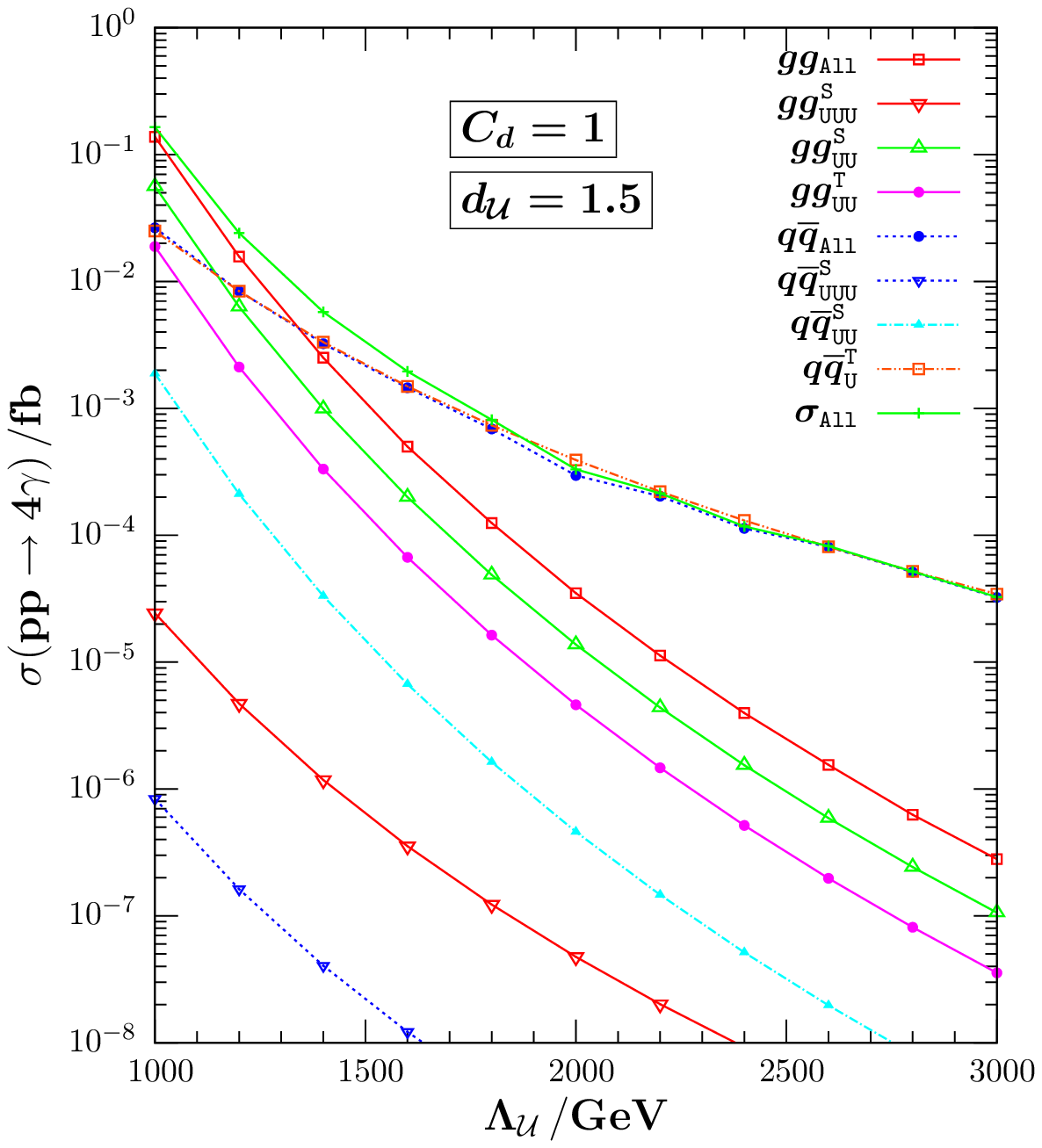}\\
	\includegraphics[width=3.5in,height=3.2in]{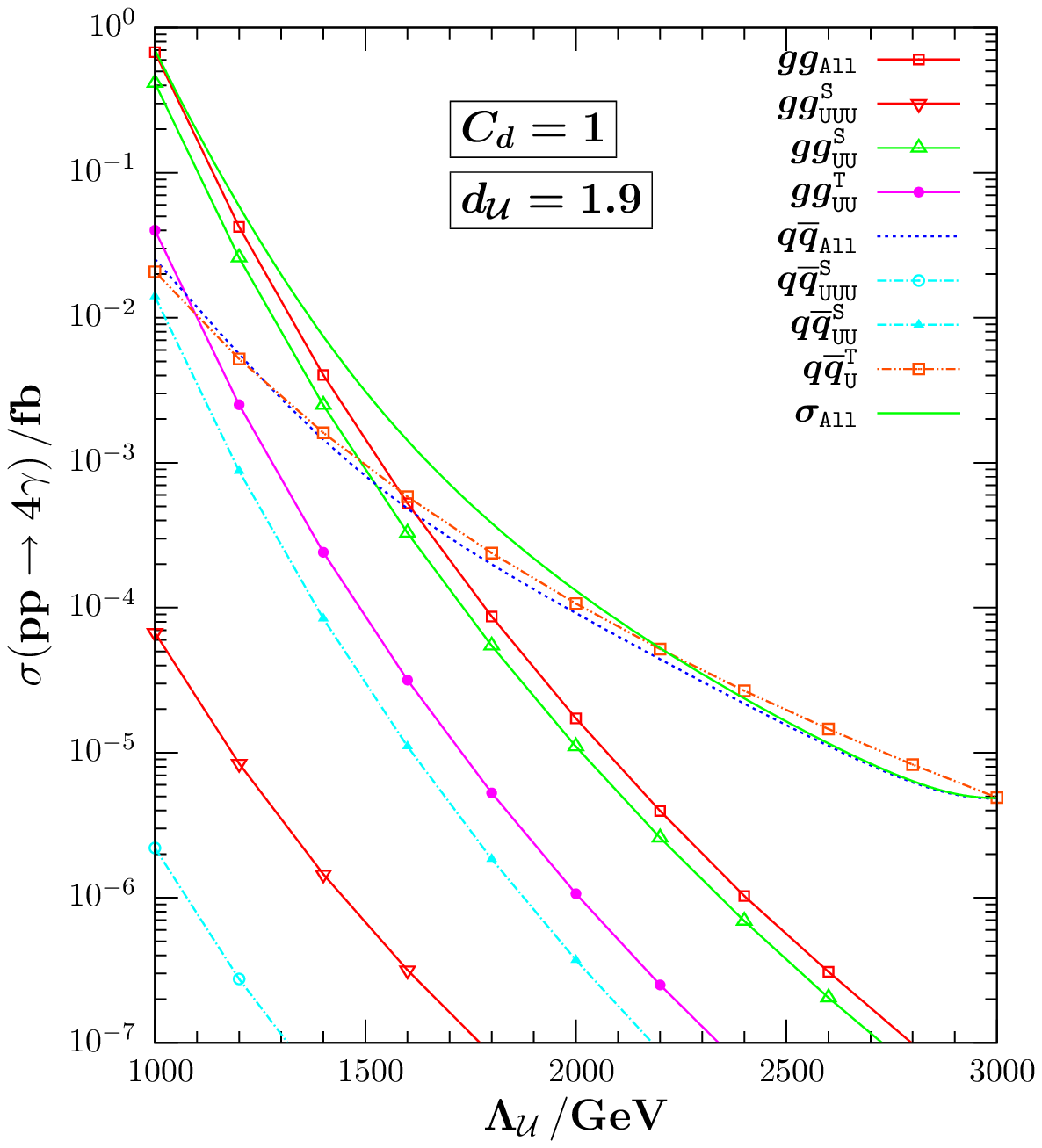}
\end{array}$
\end{center}
\vskip -0.2in
\caption{The \fourApp  cross section at $\sqrt{s}=14$ TeV as a function of $\Lambda_{\cal U}$ for $C_d=1$ and $d_{\cal U}=1.1$, $d_{\cal U}=1.5$ and $d_{\cal U}=1.9$. The individual contributions from the $gg$ and $q\bar{q}$ subprocesses are grouped and shown for each channel and number of unparticles exchanged.}\label{4Ad}
\end{figure}
\begin{figure}[htb]
\begin{center}
\hspace*{-1.2cm}
        $\begin{array}{cc}
	\includegraphics[width=3.5in,height=3.7in]{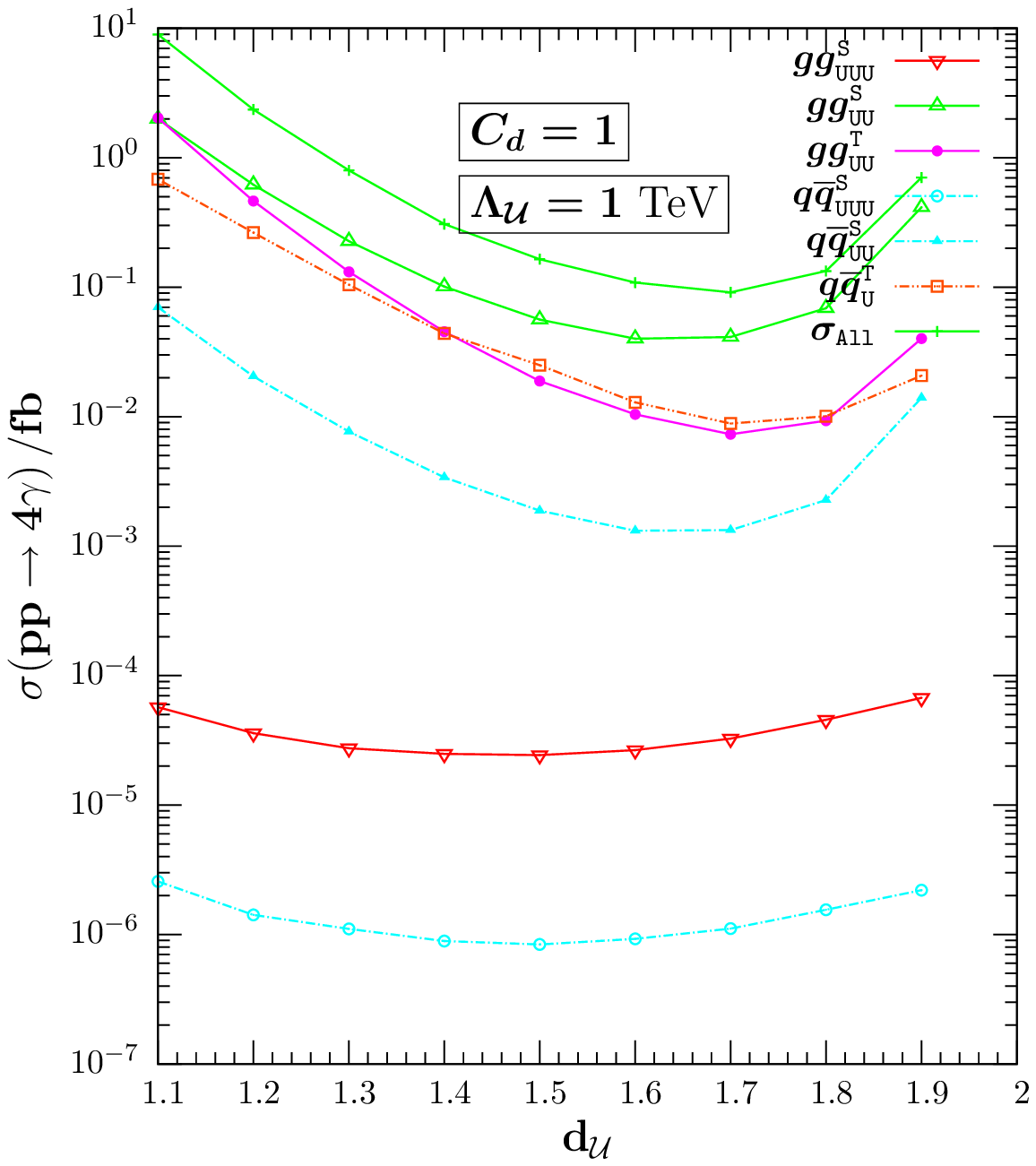} &\hspace*{-0.8cm}
	\includegraphics[width=3.5in,height=3.7in]{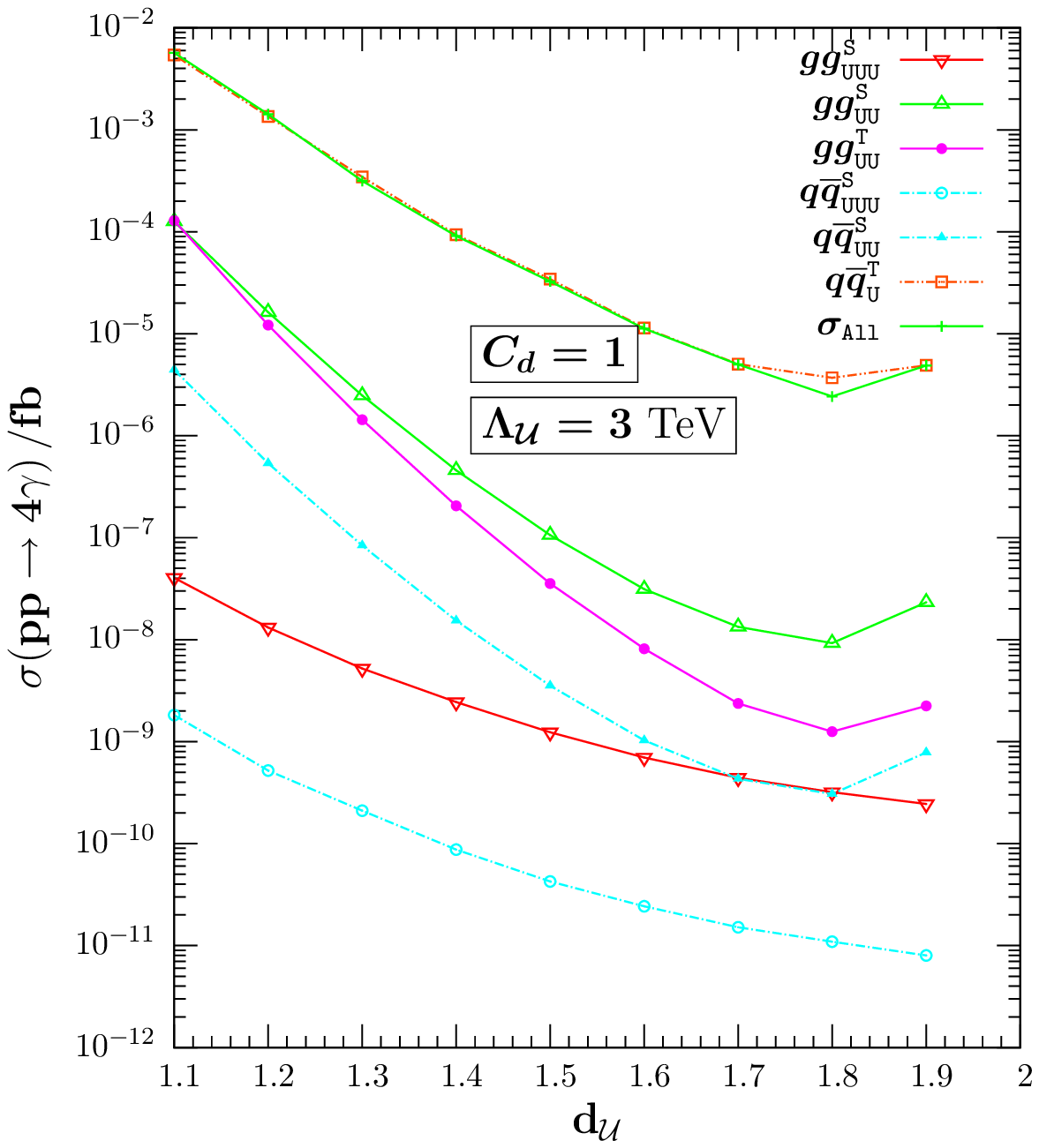}
\end{array}$
\end{center}
\vskip -0.2in
\caption{The \fourApp  cross section at $\sqrt{s}=14$ TeV as a function of $d_{\cal U}$  for $C_d=1$ and $\Lambda_{\cal U}=1$ TeV and $\Lambda_{\cal U}=3$ TeV. The individual contributions from the $gg$ and $q\bar{q}$ subprocesses are grouped and shown for each channel and number of unparticles exchanged.}\label{4Alamda}
\end{figure}

We turn our attention to the four photon production and two photon plus two gluon at the LHC and discuss these two channels separately. We take throughout the center-of-mass energy $\sqrt{s}=14$ TeV. We expect our signals to be significantly enhanced here, and even more relevant  compared to the Tevatron results if the unparticle scale $\Lambda_{\cal U}$ turns out to be large.

\subsubsection{The process \fourApp at LHC}
At the LHC,  \fourApp can proceed through $gg$ and $q{\overline q}$ at the partonic level. At $\sqrt{s}=14$ TeV the gluonic components of the proton dominate, so one would expect the cross section to be dominated by the $gg$ contribution. This is true only for a very small part of the parameter space, $\Lambda_{\cal U} <1.2$ TeV, and the result is dependent of the value of $d_{\cal U}$. (For instance, for $d_{\cal U}= 1.5$, the $q{\overline q}$ contribution dominates over the whole parameter space, whereas for $d_{\cal U}=1.1$ and $1.9$ there is small region of the parameter space where the $gg$ contribution is slightly larger.) The reason is that the $q{\overline q}$ partonic cross section can proceed through a single unparticle, while the $gg$ requires at least two.  The energy scale suppression associated with each unparticle overwhelms the advantage from the partonic distribution function of gluons versus quarks in the proton, and the correlation between the number of unparticles in a graph and the size of its contribution to the cross section dominates the cross section. Here as before there is also an indication of $s$-channel suppression, thus the smallest contribution will be given by the process with three unparticles going through the $s$ channel, with (in that instance) $gg$ contribution dominating the $q{\overline q}$ one.

In Fig.~\ref{4Alamda} we recover the expected $\Lambda_{\cal U} $ suppression (three orders of magnitude going from $\Lambda_{\cal U}=1 $ TeV to $\Lambda_{\cal U}=3 $ TeV). We also note that the total relative contribution coming from different channels is changing: at $\Lambda_{\cal U}= 1$ TeV  mainly $gg$ $s$ and $t$ channels with two unparticle exchanges dominate, and the $q{\overline q}$ $t$-channel with one unparticle exchange contributes significantly, while  at $\Lambda_{\cal U}=3 $ TeV the cross section is dominated by the one-unparticle  $q{\overline q}$ $t$-channel. This is indeed expected since larger values of $\Lambda_{\cal U}$ suppress further the channels with many unparticle exchanges. The cross section in \fourApp can reach about 10~fb, for $\Lambda_{\cal U}= 1$ TeV and $d_{\cal U}=1.1$. (Again to be expected, as  the suppression associated  with each unparticle is inversely proportional to $\Lambda_{\cal U}^{d_{\cal U}}$.)


\subsubsection{The process \twoApp at LHC}

\begin{figure}[htb]
\begin{center}
\hspace*{-1.2cm}
        $\begin{array}{cc}
	\includegraphics[width=3.5in,height=3.2in]{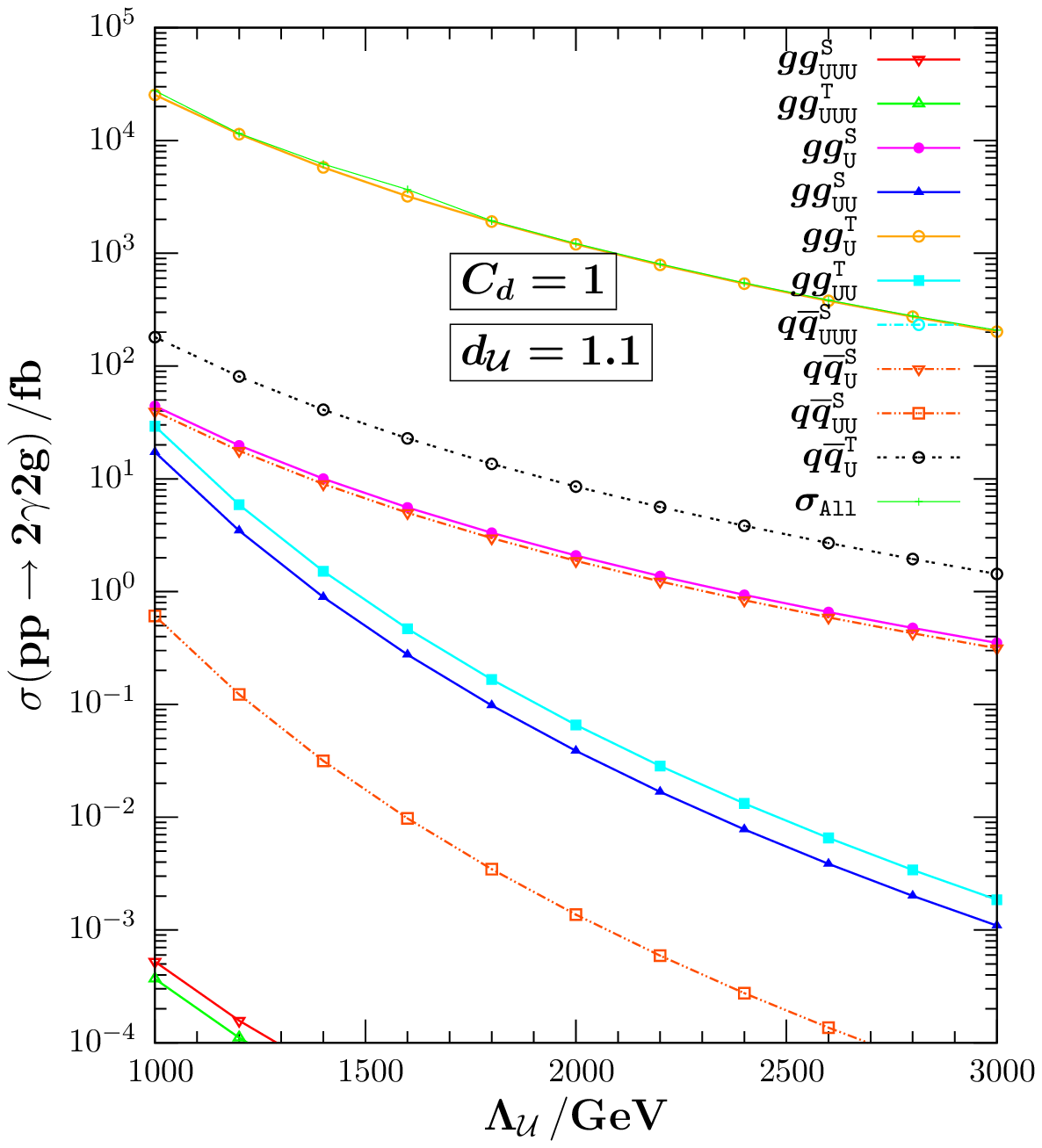} &\hspace*{-0.8cm}
	\includegraphics[width=3.5in,height=3.2in]{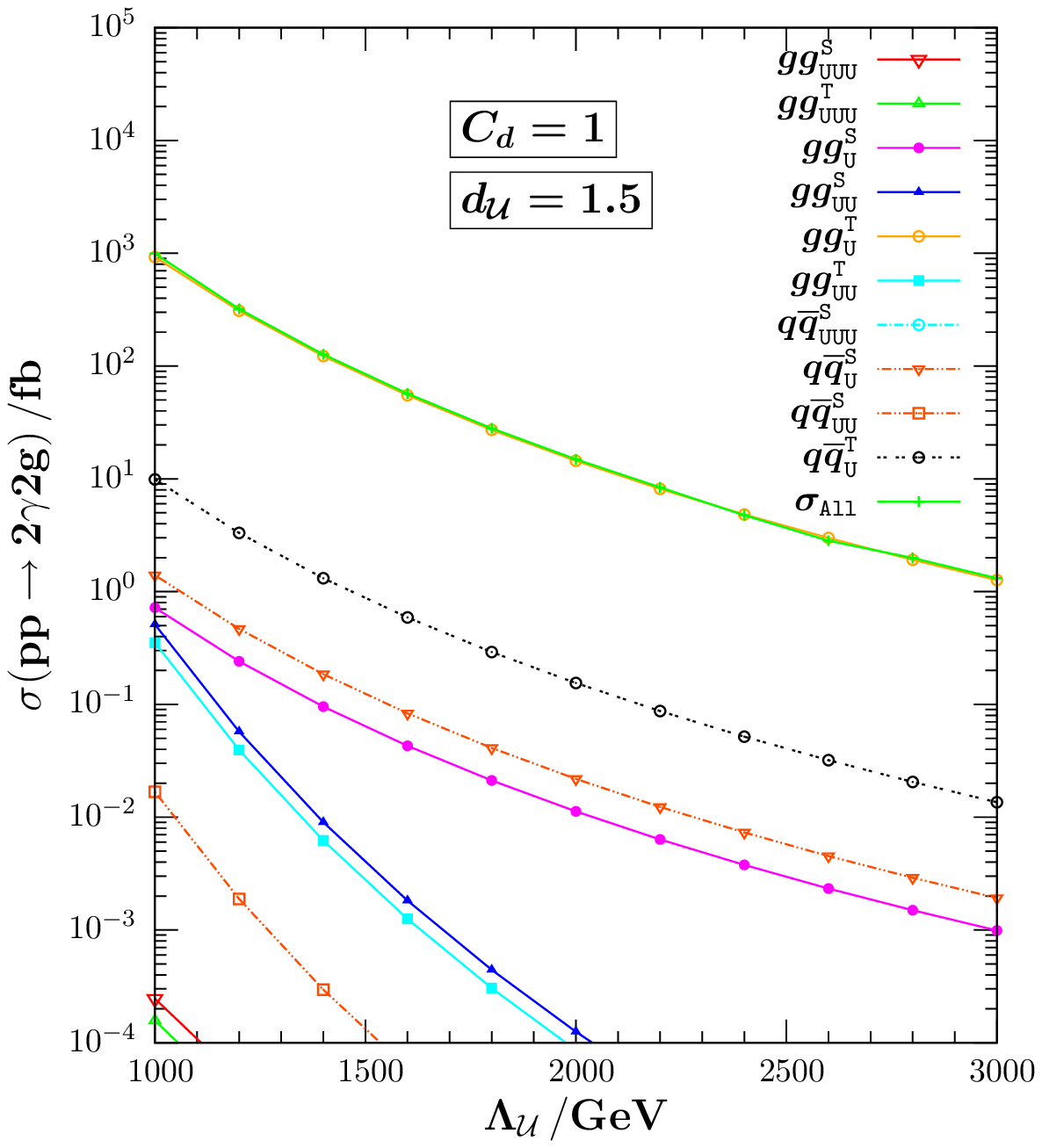}\\
	\includegraphics[width=3.5in,height=3.2in]{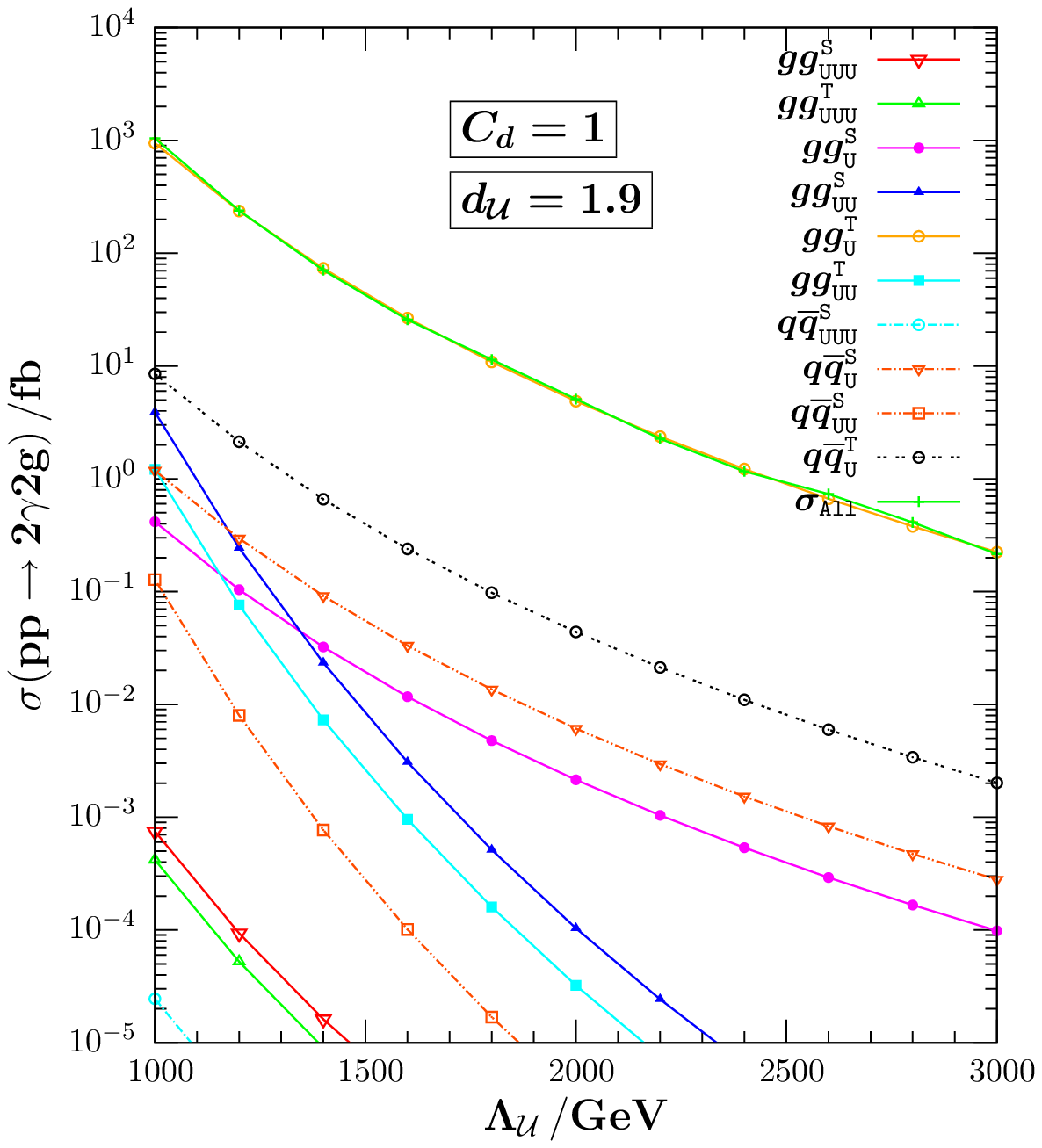}
\end{array}$
\end{center}
\vskip -0.2in
\caption{The \twoApp  cross section at $\sqrt{s}=14$ TeV as a function of $\Lambda_{\cal U}$ for $C_d=1$ and $d_{\cal U}=1.1$, $d_{\cal U}=1.5$ and $d_{\cal U}=1.9$. The individual contributions from the $gg$ and $q\bar{q}$ subprocesses are grouped and shown for each channel and number of unparticles exchanged.}\label{2A2gd}
\end{figure}
\begin{figure}[htb]
\begin{center}
\hspace*{-1.2cm}
        $\begin{array}{cc}
	\includegraphics[width=3.5in,height=3.7in]{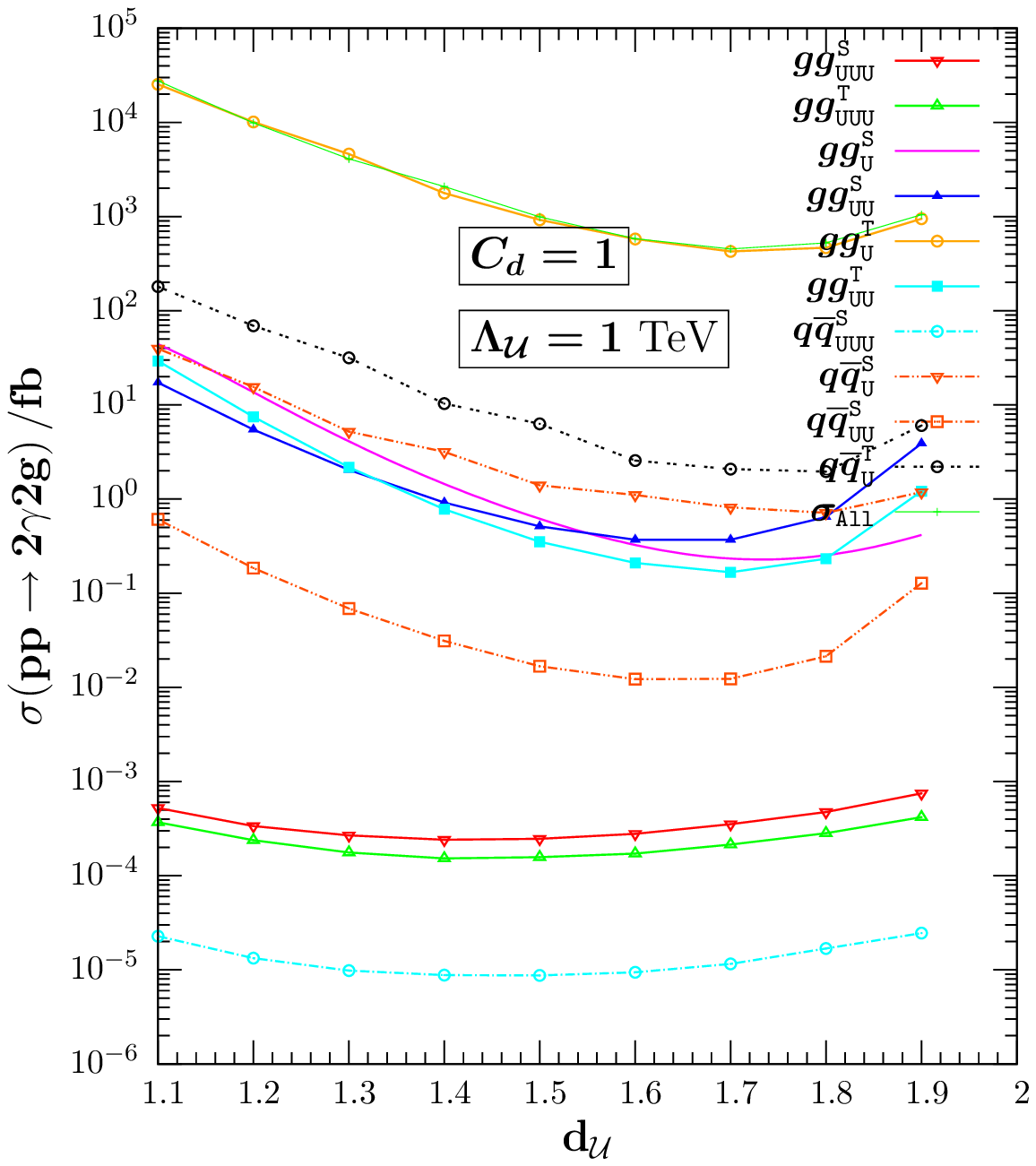} &\hspace*{-0.8cm}
	\includegraphics[width=3.5in,height=3.7in]{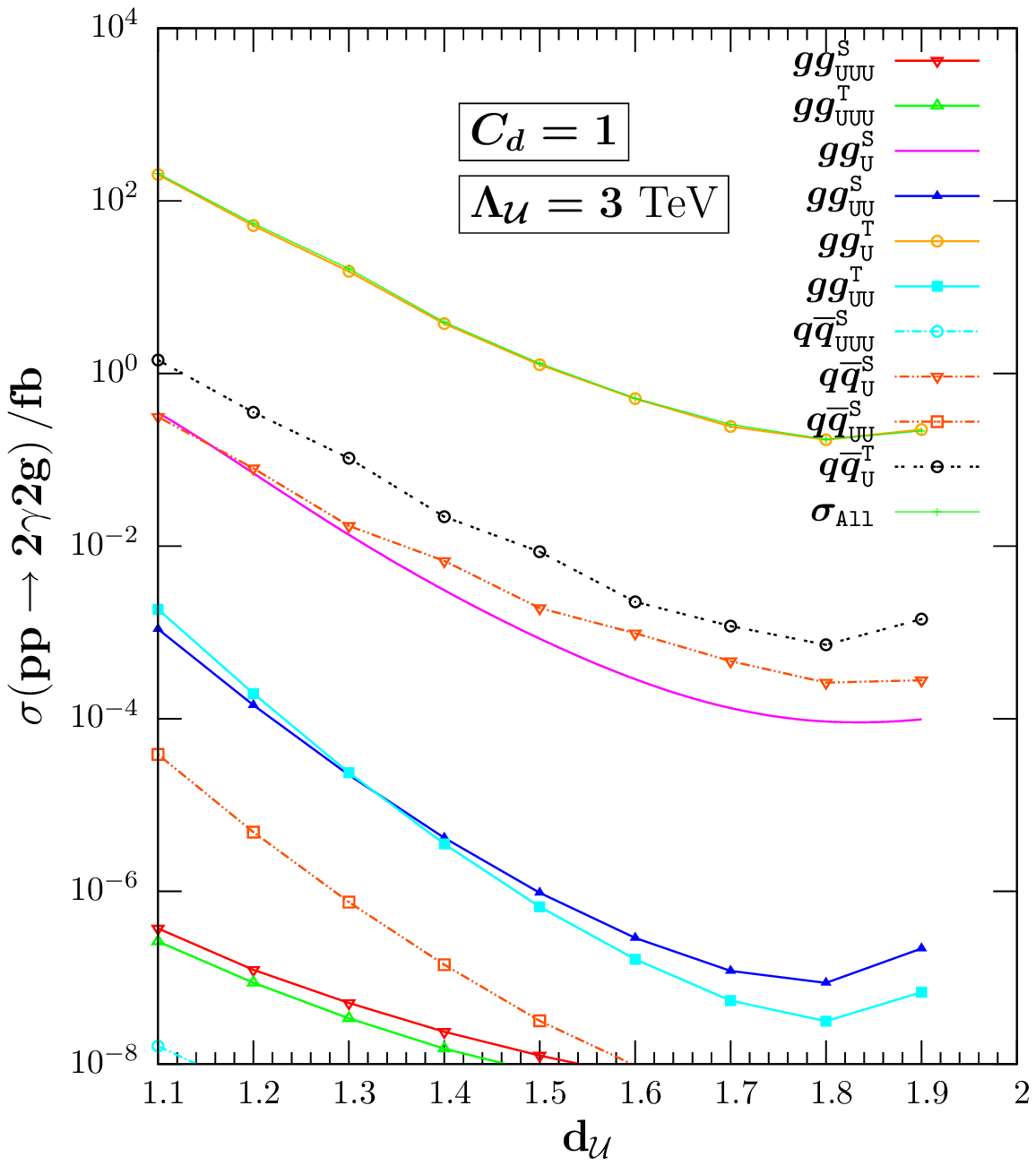}
\end{array}$
\end{center}
\vskip -0.2in
\caption{The \twoApp  cross section at $\sqrt{s}=14$ TeV as a function of $d_{\cal U}$  for $C_d=1$ and $\Lambda_{\cal U}=1$ TeV and $\Lambda_{\cal U}=3$ TeV. The individual contributions from the $gg$ and $q\bar{q}$ subprocesses are grouped and shown for each channel and number of unparticles exchanged.}\label{2A2glamda}
\end{figure}

Finally, in Figs.~\ref{2A2gd} and \ref{2A2glamda} we show the dependence of the  \twoApp cross section with the energy scale $\Lambda_{\cal U}$ and $d_{\cal U}$ parameter, respectively, at the LHC. The cross sections obtained in this case are the largest shown in this analysis, and can reach a spectacular $20-25$~pb for the most advantageous parameter point $\Lambda_{\cal U}=1$ TeV and $d_{\cal U}=1.1$. The cross sections are dominated by the $gg$ $t$-channel partonic contribution with only one unparticle in the process, although perhaps slightly less so for $d_{\cal U}=1.9$. Increasing $\Lambda_{\cal U}$ from $1$ to $3$ TeV decreases the cross section again by two orders of magnitude. It also slightly re-orders the relative  contributions coming from different channels, but not significantly. 

\subsection{The Tevatron bound on $C_d$}
\begin{figure}[b]
\begin{center}
\hspace*{-0.7cm}
        $\begin{array}{cc}
	\includegraphics[width=3.4in,height=3.7in]{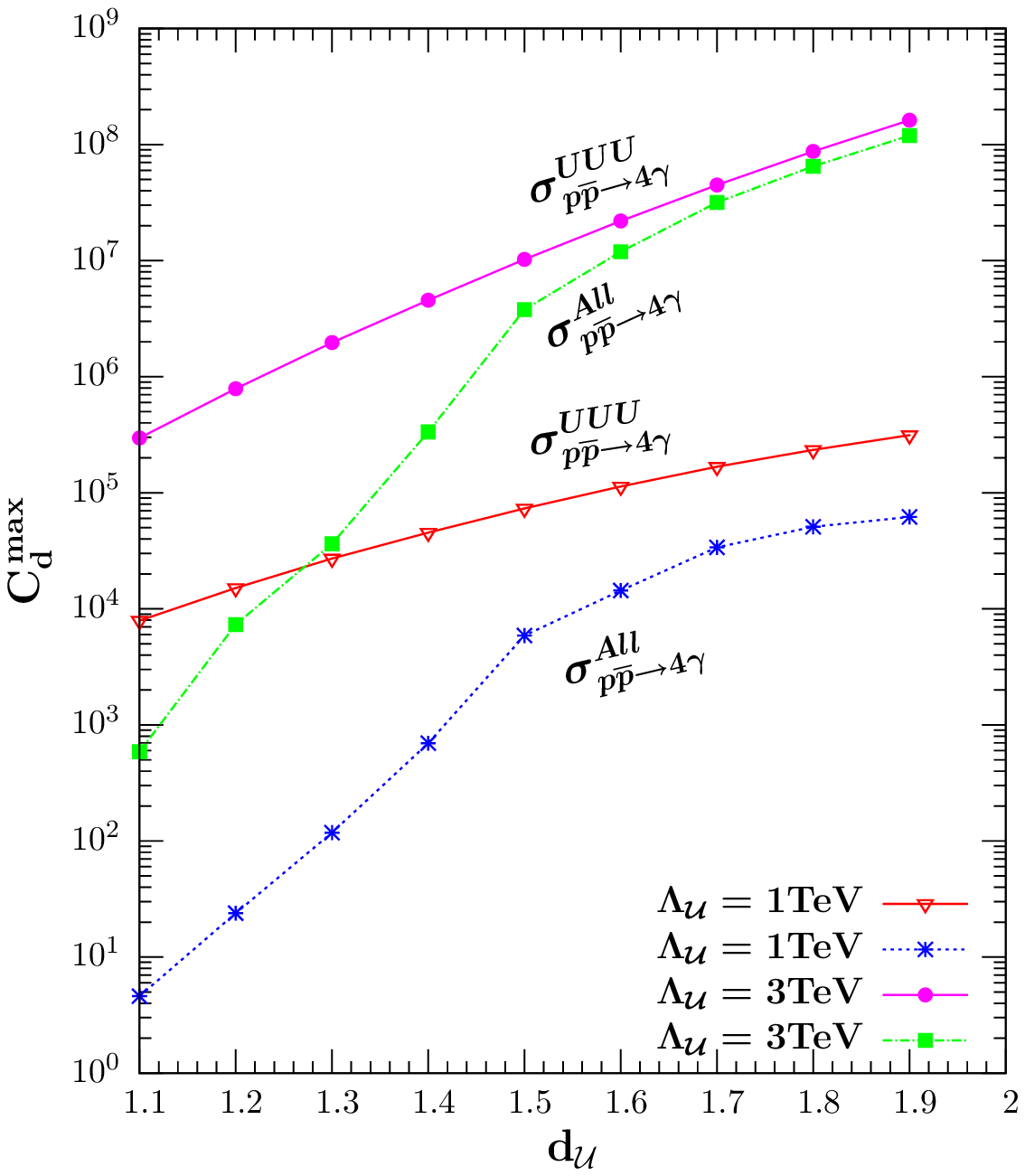} &\hspace*{-0.3cm}
	\includegraphics[width=3.4in,height=3.7in]{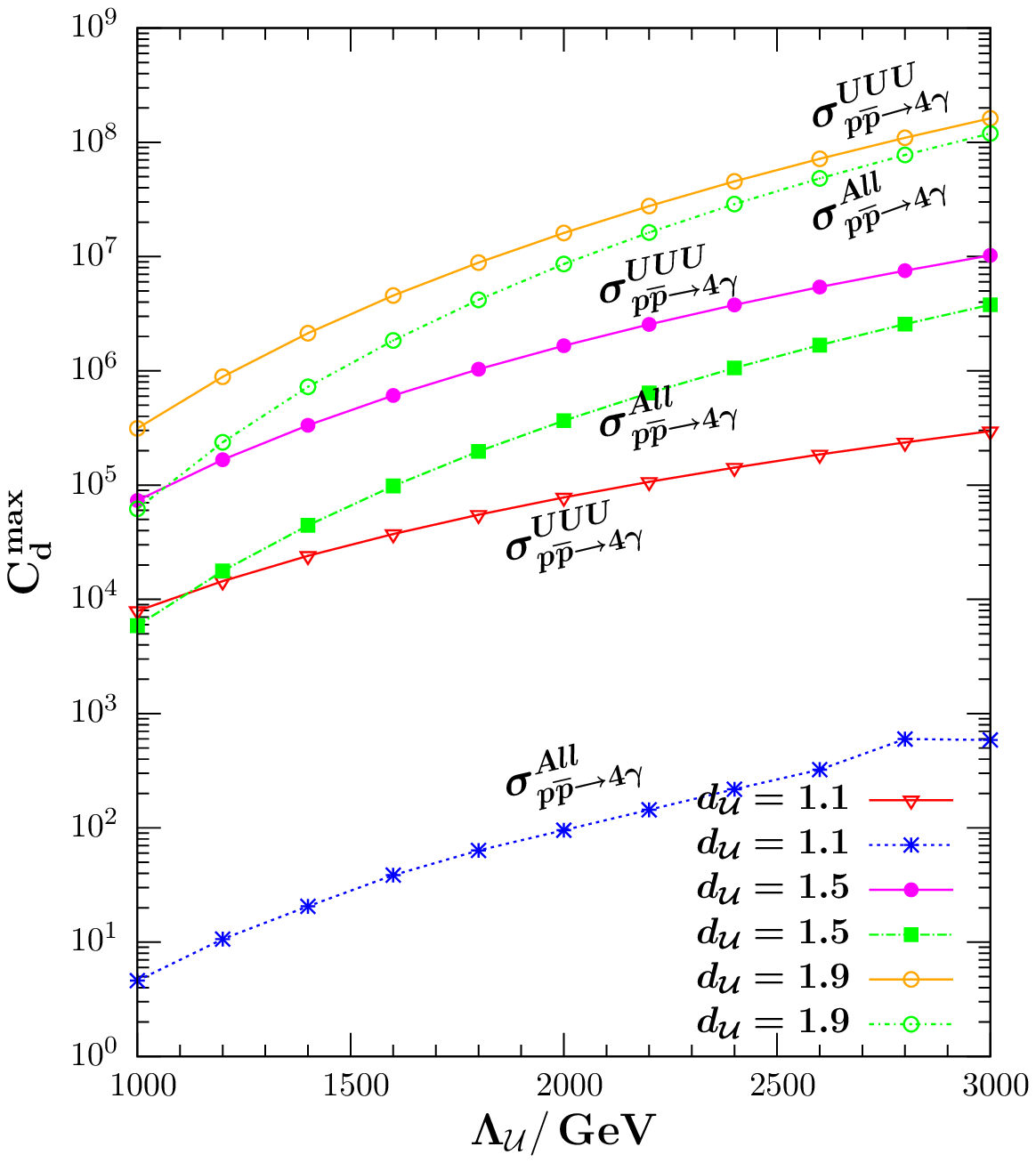}
\end{array}$
\end{center}
\vskip -0.2in
\caption{The maximal $C_d$ bounds from the Tevatron as a function of $d_{\cal U}$ for various $\Lambda_{\cal U}$ on the left panel and as a function of $\Lambda_{\cal U}$ for various $d_{\cal U}$  on the right panel. The bounds are shown separately when only the three-point interactions are considered for comparison. $\lambda_{g,\gamma}^0=1$ and $\lambda_0^\prime=\sqrt{2\pi}/e$ are used.}\label{cd_bounds}
\end{figure}

So far we set the three-point unparticle coupling constant $C_d$ as unity. However, this parameter is practically unconstrained. In Ref.~\cite{Feng:2008ae} an upper bound is obtained from the contribution of the three point interactions using the multiphoton events in the Tevatron data with integrated luminosity ${\cal L}=0.83fb^{-1}$.  Based on our results in previous subsections, we recalculate here the upper bound on $C_d$ by including all contributions to multiphoton signal. The Standard Model predicts negligible background events, and no signal event is observed above the Standard Model prediction. Then at 95\% CL the bound is \cite{Feng:2008ae}
\begin{eqnarray}
{\cal L}\times \sigma_{p\overline{p}\to 4\gamma} \le 3.04
\end{eqnarray}
where the right-hand side is the number of qualified events in the data. Here $\sigma_{p\overline{p}\to 4\gamma} \equiv \sigma_{p\overline{p}\to 4\gamma}^{q\bar{q}} + \sigma_{p\overline{p}\to 4\gamma}^{gg}$. Writing the total cross section as a sum of terms grouped in powers of $C_d$, we replace the cross section $\sigma_{p\overline{p}\to 4\gamma}$ in the left-hand side of the above expression by an quadratic equation
\begin{eqnarray}
\sigma_{p\overline{p}\to 4\gamma}^{\cal UUU}\, C_d^2 + \sigma_{p\overline{p}\to 4\gamma}^{mix}\, C_d +\left (\sigma_{p\overline{p}\to 4\gamma}^{No-{\cal UUU}}- \frac{3.04}{\cal L}\right) \le 0
\label{quad_C_d}
\end{eqnarray}
The coefficients of various $C_d$ terms are the corresponding cross sections calculated with $C_d=1$ in the preceeding subsections. Note that we cannot extract the $\Lambda_{\cal U}$ factors  from the terms in the left-hand side of Eq.~(\ref{quad_C_d}), since each scales differently with $\Lambda_{\cal U}$. So, the upper bound on $C_d$ can now be computed as the positive root of the above equation from our numerical study. The results are shown in Fig.~\ref{cd_bounds}. We separately include the results for the three-point interactions for comparison. Note that our cross section values from the three-point interactions are about a factor of 2 greater than the ones given in \cite{Feng:2008ae}, which may be due to different set of cuts used in each case as well as numerical precision in the  computation of the function $F_y$, where we had to keep the number of iterations rather low. It is seen that, in comparison to the bounds from the three-point interactions only, the upper bounds on $C_d$ are significantly reduced when all contributions are included. The deviation is much bigger for smaller values of $d_{\cal U}$. The order of reduction ranges from an order of magnitude to more than three orders of magnitude, depending on the values of $d_{\cal U}$ and $\Lambda_{\cal U}$. We obtain much stronger bounds for smaller $d_{\cal U}$ and $\Lambda_{\cal U}$ parameters region. For example, the upper bound on $C_d$ becomes around 4 at $d_{\cal U}=1.1$ and $\Lambda_{\cal U}=1$ TeV while 
it can reach $8\times 10^3$ if only the three-point unparticle interaction cross section is kept. Thus, our earlier choice of $C_d=1$ turns out to be almost the maximal value at $(d_{\cal U},\Lambda_{\cal U})=(1.1,1\rm TeV)$ and should be considered conservative for any set of bigger $(d_{\cal U},\Lambda_{\cal U})$ values.

Now, what would be the effect of relaxing $C_d=1$ and using maximal $C_d$ on the processes $pp(\overline{p}) \to  \gamma \gamma \gamma \gamma$ and $pp(\overline{p}) \to  \gamma \gamma gg$ at both colliders? First of all,  this would not  change the total cross section significantly, as long as the cross section is dominated by the single unparticle exchange in either $s$ or $t$ channel. Only if the interference terms between the three-point interactions and the rest of the interactions to the cross section are sizable,  will there be such enhancement, which is linear in $C_d^{max}$. Of course, if we focus on the the case of individual contributions enhancement, the three-point unparticle interaction part will receive the largest enhancement since it scales as $C_d^{max}$ while in the rest there will be no change. A more precise analysis would first fit the curves in Fig.~\ref{cd_bounds} to get an approximate relation for $C_d^{max}$ as function of $d_{\cal U}$ when $\Lambda_{\cal U}$ is fixed. Then a  search for maximal value of the cross section would use these functions to also obtain the relative contributions. We find it unnatural to go very large $C_d$ values, as the bounds on it most likely will become stronger once some data from LHC is avaliable. As we did not include this in our evaluations, our total cross section values obtained are  more conservative.

\section{Summary and Conclusion}\label{sec:conclusion}
Since the original suggestion of Georgi \cite{Georgi:2007ek}, and the formulation of unparticle physics, viewed as a conformal theory with exact scale invariance coupled to the standard model at high energies,  numerous phenomenological explorations of this idea have been developed.

In the present work, we have presented a contribution to the study of implications of self-interactions of the low energy conformal sector. We concentrated on implications of such couplings on the multiphoton signal at the hadronic colliders, the Tevatron and the LHC. The processes $pp(\overline{p}) \to \gamma \gamma\gamma \gamma$ and $\gamma \gamma gg$ involve exchanges of one, two and three unparticles as intermediate states. The diagrams contributing to such processes involve unparticle exchanges, as well as unparticle and vector boson exchanges. The situation is unusual, and evaluation of these diagrams does not resemble expressions obtained form ordinary perturbation theory, as graphs with one, two or three internal unparticle exchanges contribute to the same subprocess. As every unparticle vertex contains a factor of $\displaystyle \frac{\lambda_{g, \gamma}^0}{\Lambda_{\cal U}^{d_{\cal U}}}$, while every SM particle couples to gluons (photons) with interaction strength $\alpha_s(\alpha_w)$, the evaluation of the cross section includes a sum over terms with different powers of coupling constants and of unparticle energy scale $\Lambda_{\cal U}$. Although this process is suppressed (in fact negligible) in the SM and can only proceed here though (at least one) unparticle exchange, the total cross section is dominated by the diagrams with the minimal number of internal unparticles. We have calculated every contribution to the cross section separately, discriminating between $s$ and $t$ channels, number of internal unparticles, and partonic components. We confirm earlier results for the contribution of the three  unparticle exchange and correlation function, but also show that this contribution is subdominant.

At the Tevatron and the LHC, multiphoton signals coming from unparticle exchanges give rise to significant cross sections, the most spectacular of which could be as large as $20-25$~pb, in  \twoApp at LHC, for the most advantageous parameter point, the minimum $\Lambda_{\cal U}$ and $d_{\cal U}$ investigated ($\Lambda_{\cal U}=1$ TeV and $d_{\cal U}=1.1$). We consider first the case in which $C_d$, the coupling from the three-unparticle correlation function is set to one. We then  restrict the $C_d$ parameter based on the Tevatron data. This was done previously, assuming the multiphoton process is dominated entirely by the three-unparticle exchange, where expressions for $C_d$ can be  obtained independently of the unparticle scale $\Lambda_{\cal U}$. In a complete evaluation, terms with different numbers of unparticles enter the summation, and the restriction on $C_d$ becomes a restriction on the quadratic roots of a $\Lambda_{\cal U}$-dependent equation, rather than a simple inequality. For completness, we present maximal bounds on $C_d$ for both three-unparticle exchanges, as well as for the total cross section.

In concluding, the main point of our paper is to show that multiphoton signals at hadron collider are significant if one includes the coupling of SM particles with one or several unparticles, and if observed, these events  would be a strong indication of   unparticle physics.

\section{Acknowledgments}
The work of M.F. and I.T. was supported in part by NSERC of
Canada under the Grant No. SAP01105354.  We thank Thomas Hahn for many helpful hints for customization of the softwares {\tt FeynArts} and {\tt FormCalc}.

\section{Appendix}

In this section we give all the relevant Feynman diagrams for the processes studied. We include only diagrams in which at least one scalar unparticle ${\cal U}_s$ contributes. The Standard Model diagrams are not included.

\subsection{Feynman Diagrams for the process $pp(\overline{p}) \to  \gamma \gamma \gamma \gamma $}
\begin{figure}[htb]
\vskip 0.2in
\begin{center}
        $\begin{array}{c}
	\includegraphics[width=2.8in]{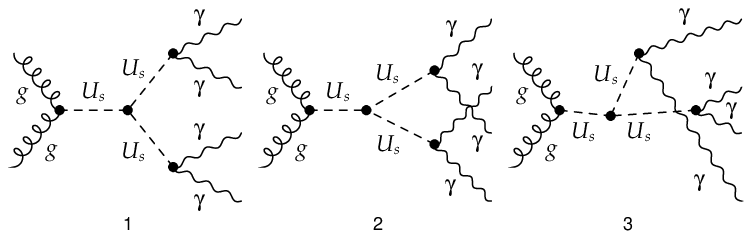}\\
        (a)
        \\
	\\
	\includegraphics[width=5.5in]{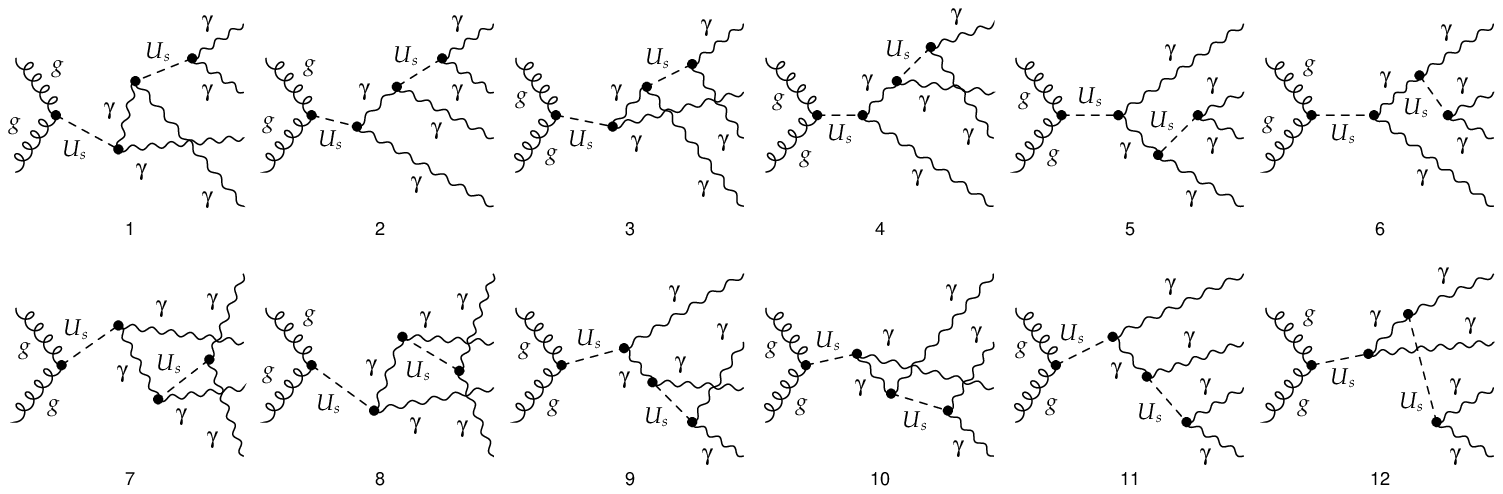}\\
        (b)
        \\
	\\
	\includegraphics[width=5.5in]{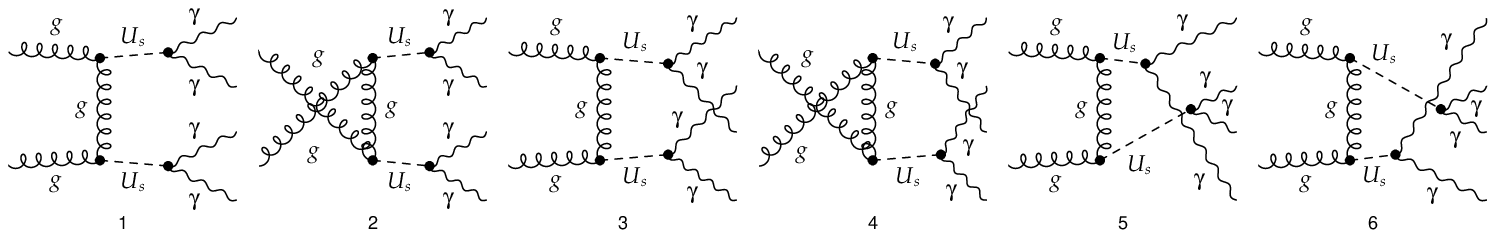}\\
        (c)
\end{array}$
\end{center}
\vskip -0.2in
\caption{The Feynman diagrams contributing to the subprocess \fourAgg  with the three-point unparticle vertex (panel $a$), the s-channel (panel $b$) and the t- and u-channels (panel $c$). } \label{fd_gg4A}
\end{figure}

\begin{figure}[htb]
\vskip 0.2in
\begin{center}
        $\begin{array}{c}
	\includegraphics[width=2.8in]{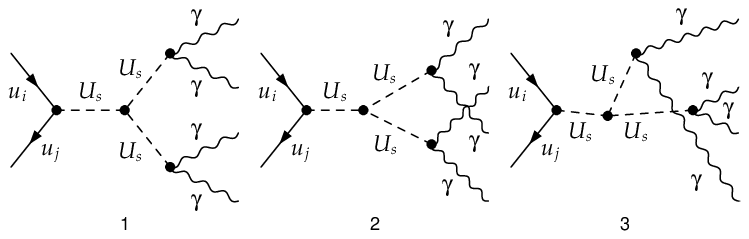}\\
        (a)
        \\
	\\
	\includegraphics[width=5.5in]{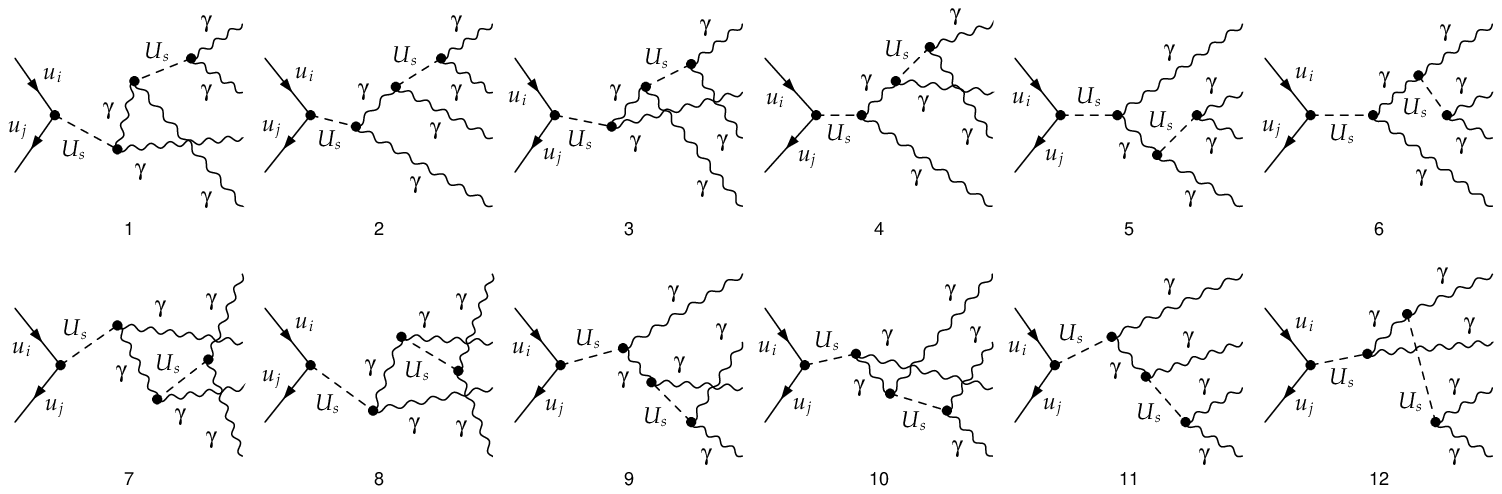}\\
        (b)
        \\
	\\
	\includegraphics[width=5.5in]{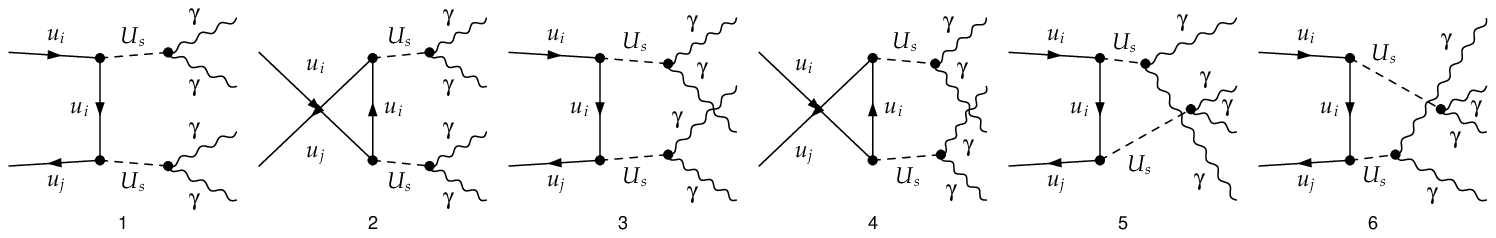}\\
        (c)
\end{array}$
\end{center}
\vskip -0.2in
\caption{The Feynman diagrams contributing to the subprocess \fourAqq  with the three-point unparticle vertex (panel $a$), the s-channel (panel $b$) and in the t- and u-channels with two unparticle ${\cal U}_s$ exchanged (panel $c$). The quark $q_{i,j}$ represent all five light quarks and $i=j$.}\label{fd_qq4A_1}
\end{figure}

\begin{figure}[htb]
\begin{center}
	\includegraphics[width=5.2in]{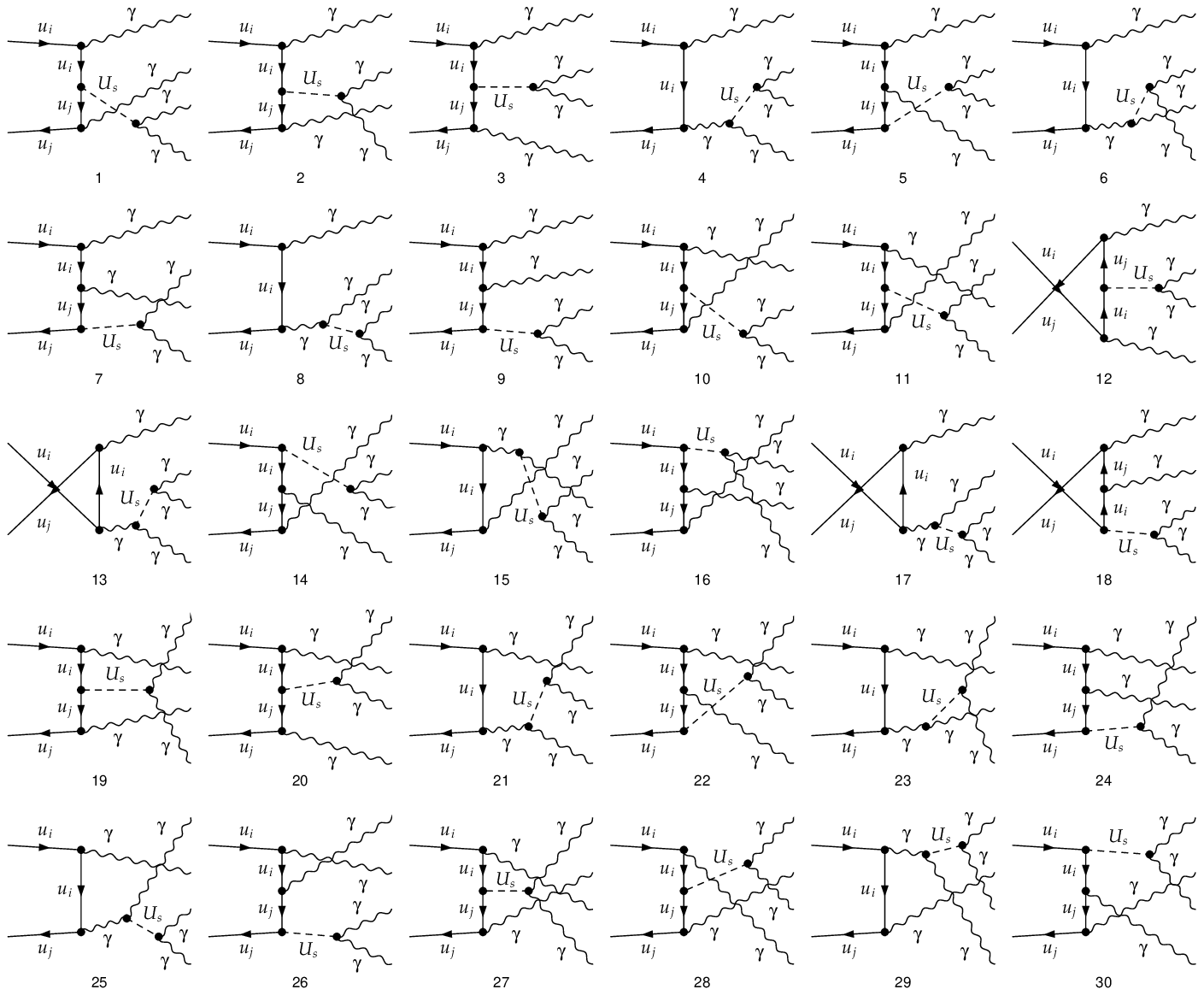}
%
	\includegraphics[width=5.2in]{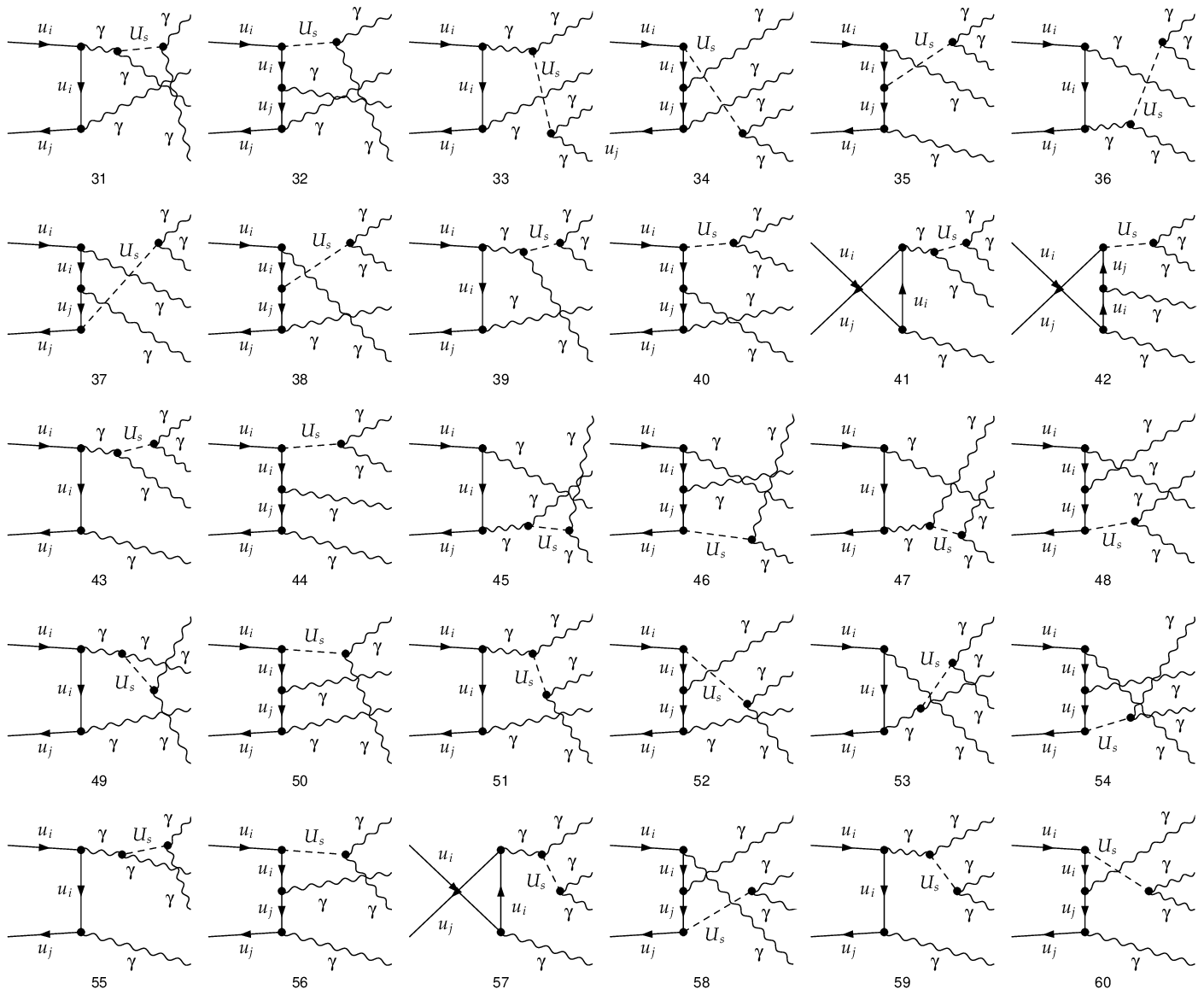}
\end{center}
\vskip -0.2in
\caption{The Feynman diagrams contributing to the subprocess \fourAqq in the t-channel with a single unparticle ${\cal U}_s$ exchanged. The corresponding u-channel diagrams are also included.}\label{fd_qq4A_2}
\end{figure}

We start first with the $pp(\overline{p}) \to  \gamma \gamma \gamma \gamma $ and group the figures  into two: one for the $gg$ partonic contribution, and the other for $q {\overline q}$ partonic contributions. Following the numerical estimates, we group further together graphs with the same number of unparticles in the intermediate states, and the processes going through the $s$ and $t-u$ channels together. They are all given in Figs.~\ref{fd_gg4A}, \ref{fd_qq4A_1}, and \ref{fd_qq4A_2}. In all the diagrams, the wave and the spiral lines represent photon and gluon fields, respectively.



\thispagestyle{empty}

\clearpage
\subsection{Feynman Diagrams for the process $pp(\overline{p}) \to  \gamma \gamma gg$}
\begin{figure}[htb]
\vskip 0.2in
\begin{center}
        $\begin{array}{cc}
	\hspace*{-8cm} \includegraphics[width=1.0in]{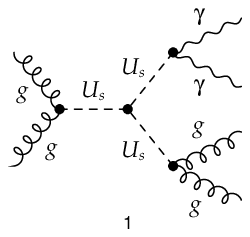} &\hspace*{-8cm} \includegraphics[width=1.8in]{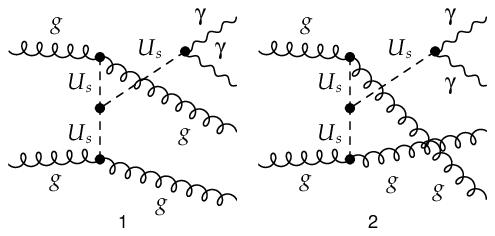}\\
        \hspace*{-0.5cm}(a)\hspace*{7cm}  (b)
	\\
	\\
	\includegraphics[width=5.5in]{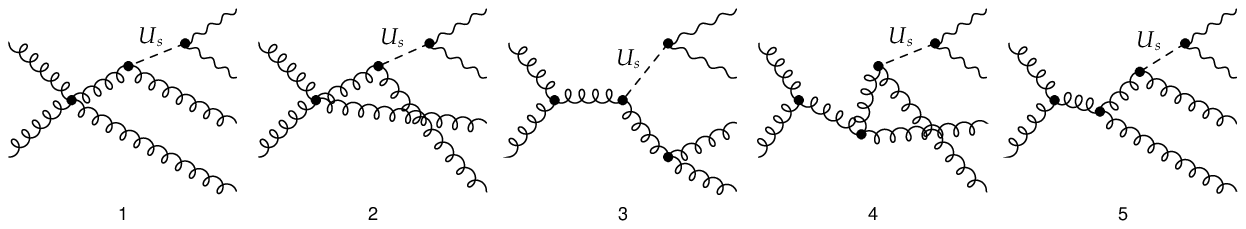} & \\
        (c)
	\\
	\\
	\includegraphics[width=5.0in]{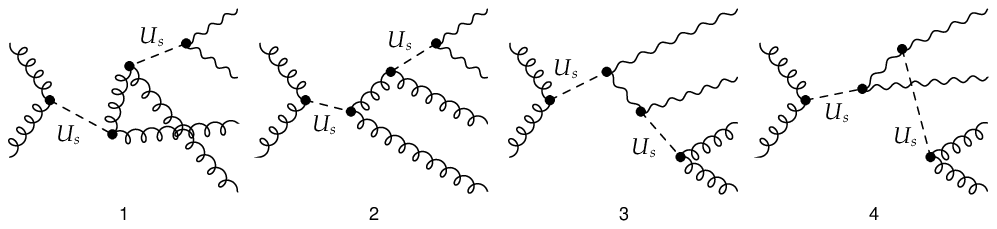} & \\
        (d)
\end{array}$
\end{center}
\vskip -0.2in
\caption{The Feynman diagrams contributing to the subprocess \twoAgg  with the three-point unparticle vertex in the s-channel (panel $a$), the three-point unparticle vertex in the t-channel (panel $b$), the s-channel with single unparticle exchange (panel $c$) and the s-channel with two unparticle exchange (panel $d$).}\label{fd_gg2A2g_1}
\end{figure}

\begin{figure}[htb]
\vskip 0.2in
\begin{center}
        $\begin{array}{c}
	\includegraphics[width=5.5in]{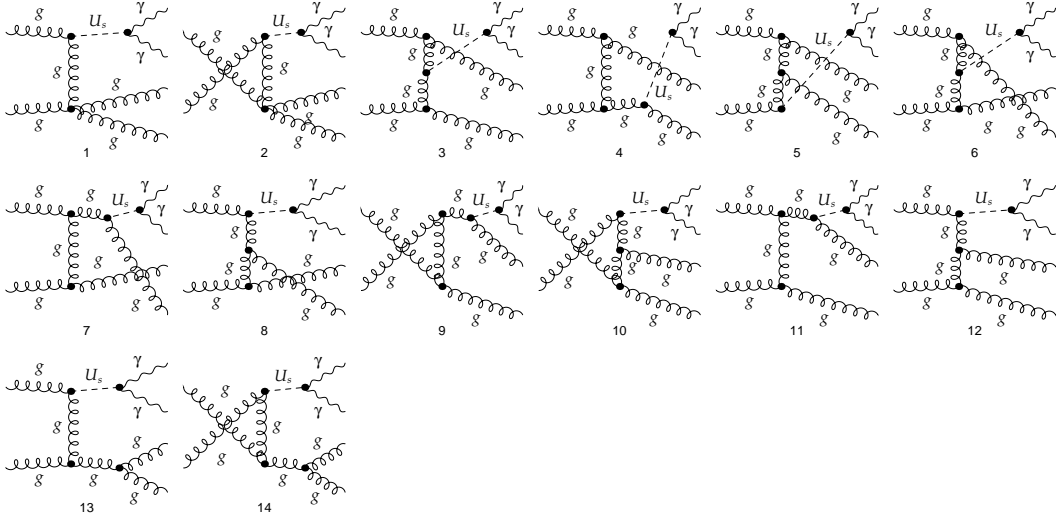}\\
        (a)
        \\
	\\
	\includegraphics[width=5.5in]{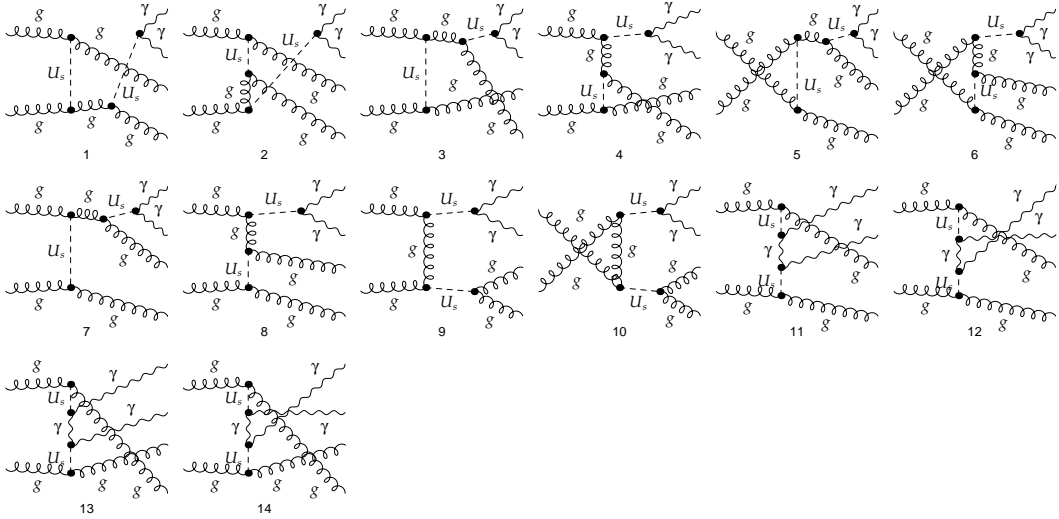}\\
        (b)
\end{array}$
\end{center}
\vskip -0.2in
\caption{The Feynman diagrams contributing to the subprocess \twoAgg in the t-channel with a single unparticle ${\cal U}_s$ exchanged (panel $a$) and in the t-channel with two unparticles ${\cal U}_s$ exchanged (panel $b$). The corresponding u-channel diagrams are also included.}\label{fd_gg2A2g_2}
\end{figure}

\begin{figure}[htb]
\vskip 0.2in
\begin{center}
        $\begin{array}{cc}
	\hspace*{-10cm} \includegraphics[width=1.0in]{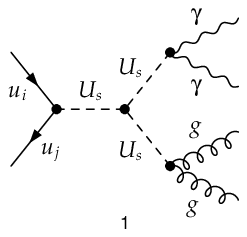} & \hspace*{-10cm}\includegraphics[width=3.0in]{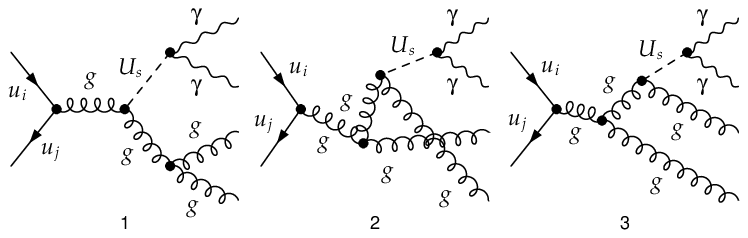} \\
       \hspace*{-2.0cm} (a) \hspace*{7.5cm} (b)
	\\
	\\
	\hspace*{-4.5cm}\includegraphics[width=4.2in]{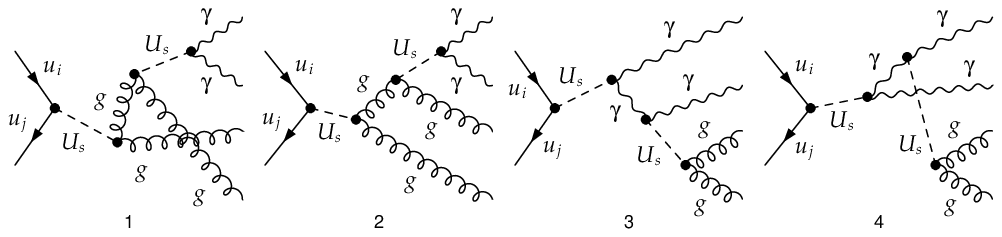} & \hspace*{-3.5cm} \includegraphics[width=2.0in]{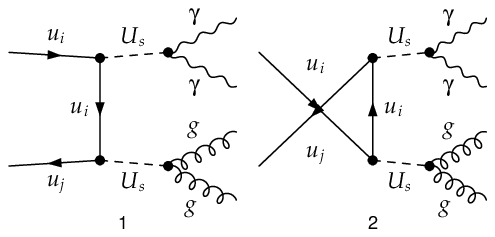} \\
        \hspace*{6cm} (c) \hspace*{8cm} (d)
\end{array}$
\end{center}
\vskip -0.2in
\caption{The Feynman diagrams contributing to the subprocess \twoAqq  with the three-point unparticle vertex (panel $a$), the s-channel with a single unparticle ${\cal U}_s$ exchanged (panel $b$) and the t-channel with two unparticles ${\cal U}_s$ exchanged (panel $c$) and  in the t- and u-channels with two unparticle ${\cal U}_s$ exchanged (panel $d$). The quark $q_{i,j}$ represent all five light quarks and $i=j$.}\label{fd_qq2A2g_1}
\end{figure}

\begin{figure}[htb]
\begin{center}
	\includegraphics[width=5.5in]{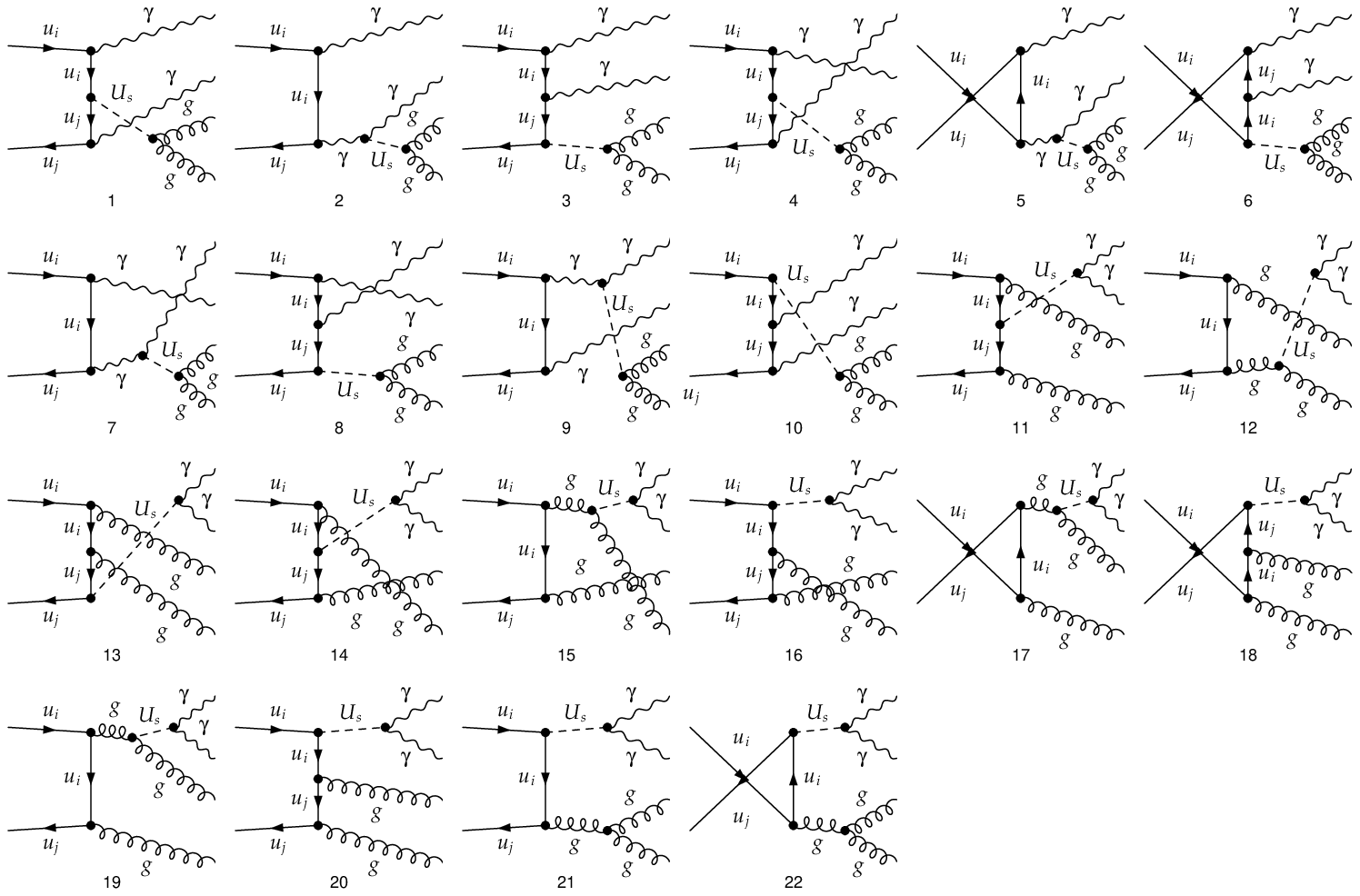}
\end{center}
\vskip -0.2in
\caption{The Feynman diagrams contributing to the subprocess \twoAgg in the t-channel with a single unparticle ${\cal U}_s$ exchanged. The corresponding u-channel diagrams are included.}\label{fd_qq2A2g_2}
\end{figure}

Next we present the Feynman graphs for the process $pp(\overline{p}) \to \gamma \gamma gg$. The organization of these is the same as for $pp(\overline{p}) \to \gamma \gamma \gamma \gamma$: we classify contributions from different partonic components first, then further group them according to the number of unparticles and the channel through which they proceed. They are all given in Figs.~\ref{fd_gg2A2g_1}, \ref{fd_gg2A2g_2}, \ref{fd_qq2A2g_1} and \ref{fd_qq2A2g_2}.


\end{document}